\documentclass{aa}
\usepackage{txfonts,graphicx}
\usepackage{longtable,lscape}
\usepackage{natbib}
\bibpunct{(}{)}{;}{a}{}{,}

\begin{document}

\title{{\em XMM-Newton} observations of the young open cluster around
$\lambda$~Orionis
\thanks{Based on observations obtained with {\em XMM-Newton}, an ESA science
mission with instruments and contributions directly funded by ESA Member
States and NASA}}

\author{E. Franciosini\inst{1}
\and 
G.G.~Sacco\inst{2}
}

\institute{
INAF -- Osservatorio Astrofisico di Arcetri, Largo E. Fermi 5, 50125
Florence, Italy, \email{francio@arcetri.astro.it}
\and
Chester F. Carlson Center for Imaging Science, Rochester Institute of
Technology, 54 Lomb Memorial Drive, 14623 Rochester, USA}

\date{Received 21 June 2010 / Accepted 30 March 2011}
\titlerunning{{\em XMM-Newton} observations of the $\lambda$~Ori cluster}

\abstract{}{
We studied the X-ray properties of the young ($\sim$\,1$-$8~Myr) open cluster
around the hot (O8\,III) star $\lambda$~Ori and compared them with those of
the similarly-aged $\sigma$~Ori cluster to investigate possible effects of
the different ambient environment.
}{
We analysed an {{\em XMM-Newton}} observation of the cluster using EPIC
imaging and low-resolution spectral data. We studied the variability of
detected sources, and performed a spectral analysis of the brightest sources
in the field using multi-temperature models.
}{
We detected 167 X-ray sources, of which 58 are identified with known cluster
members and candidates, from massive stars down to low-mass stars with
spectral types $\sim$\,M5.5. Another 23 sources were identified with new
possible photometric candidates. Late-type stars have a median $\log
L_\mathrm{X}/L_\mathrm{bol}\sim -3.3$, close to the saturation limit.
Variability was observed in $\sim$\,35\% of late-type members or candidates,
including six flaring sources. The emission from the central hot star
$\lambda$~Ori is dominated by plasma at 0.2$-$0.3~keV, with a weaker
component at 0.7~keV, consistently with a wind origin. The coronae of
late-type stars can be described by two plasma components with temperatures
$T_1\sim\,$0.3$-$0.8~keV and $T_2\sim\,$0.8$-$3~keV, and subsolar abundances
$Z\sim$\,0.1$-$0.3\,$Z_\odot$, similar to what is found in other star-forming
regions and associations. No significant difference was observed between
stars with and without circumstellar discs, although the smallness of the
sample of stars with discs and accretion does not allow us to draw
definitive conclusions.
}{
The X-ray properties of $\lambda$~Ori late-type stars are comparable to
those of the coeval $\sigma$~Ori cluster, suggesting that stellar activity
in $\lambda$~Ori has not been significantly affected by the different
ambient environment.}


\maketitle

\section{Introduction}
\label{intro}

The \object{$\lambda$~Ori cluster} (\object{Collinder~69}), located at a
distance of about 400~pc \citep{murdin77,mayne08}, consists of a group of
$\sim$\,10~OB stars and $\sim$\,200 late-type pre-main sequence (PMS) stars
concentrated within 1\,deg of the O8\,III $+$ B0\,V binary
\object{$\lambda$~Ori\,AB}. The cluster lies at the centre of an \ion{H}{ii}
region delimited by a dense ring of molecular gas and dust with a 9\,deg
diameter \citep{mm87,zhang89}. Based on an extensive optical photometric and
medium-resolution spectroscopic survey of the entire region,
\citet{dm99,dm01,dm02} suggested that star formation in the region started
$\sim$\,6$-$8~Myr ago, and was interrupted $\sim$\,1$-$2~Myr ago by a
supernova explosion which dispersed the parent gas cloud, creating the
molecular ring, and unbound the cluster. They also found that the fraction
of classical T~Tauri stars belonging to the $\lambda$~Ori cluster was only
$\sim$\,7\%, significantly lower than other clusters and star-forming
regions (SFRs) of similar age, and suggested that circumstellar discs might
have been photoevaporated by the far-UV radiation of the hot stars before
the supernova explosion, when low-mass and OB stars were still confined by
the parent cloud in a smaller region. {\em Spitzer} imaging by
\citet{barrado07} showed that only $\sim$\,30\% of low-mass cluster members
have circumstellar discs. \citet{sacco08} compared the disc and accretion
properties of low-mass stars in $\lambda$~Ori with those of the
similarly-aged cluster $\sigma$~Ori, finding that not only the fraction of
stars with discs, but also the fraction of discs that are actively accreting
is significantly lower in $\lambda$~Ori than in $\sigma$~Ori. These authors
suggested that the observed discrepancy might be due either to the effect of
the massive stars and the supernova explosion, or to an older age of the
$\lambda$~Ori cluster with respect to $\sigma$~Ori, although no definitive
conclusion could be drawn from the available data. 

An interesting question to answer is whether the supernova explosion and the
different ambient environment might have affected the magnetic activity of
PMS stars in the $\lambda$~Ori cluster. To investigate this issue, we
performed an X-ray observation of the $\lambda$~Ori cluster using the {\em
XMM-Newton} satellite. The observation was centred on the hot star
$\lambda$~Ori\,AB, in order to obtain both a high-resolution RGS spectrum of
the central source and EPIC imaging data and low-resolution spectra over the
whole field of view. A detailed analysis of the RGS spectrum of
$\lambda$~Ori\,AB will be presented in a forthcoming paper. Here we
concentrate on the analysis of the EPIC data, to derive the X-ray properties
of the cluster population; we will then compare the results with those
obtained with {\em XMM-Newton} for the \object{$\sigma$~Ori cluster} by
\citet[hereafter FPS06]{francio06sig} to investigate possible differences
between the two clusters.

The $\lambda$~Ori region was first observed in X-rays with {\em Einstein},
which found five X-ray sources, one identified with the hot star
$\lambda$~Ori \citep{stone85}. The {\em ROSAT} All-Sky Survey detected
several X-ray sources identified with new T~Tauri stars
\citep{sterzik95,neuh95,neuh97,alcala96,magazzu97}. The central hot star was
observed with {\em ASCA} by \citep{corcoran94} who found a hard spectrum
with temperatures of 0.3 and 2.3~keV. Recently, \citet{barrado11} has
performed an {\em XMM-Newton} observation of two fields to the east and west
of the cluster centre, partially overlapping our observation. However, a
detailed X-ray study of the central region of the cluster has not been done
before. Our observations represent the first comprehensive analysis of the
X-ray properties of the low-mass population located around the hot star
$\lambda$~Ori.

The paper is organised as follows. The X-ray observations and data analysis
are described in Sect.~\ref{observ}. In Sects.~\ref{var} and \ref{spectra}
we present the results of the variability and spectral analysis, while in
Sect.~\ref{xlum} we discuss the X-ray luminosities of cluster members and
candidates. In Sect.~\ref{sigori} we compare our results with those obtained
for the $\sigma$~Ori cluster. Conclusions are given in Sect.~\ref{concl}.

\begin{figure}
\resizebox{\hsize}{!}{\includegraphics[clip]{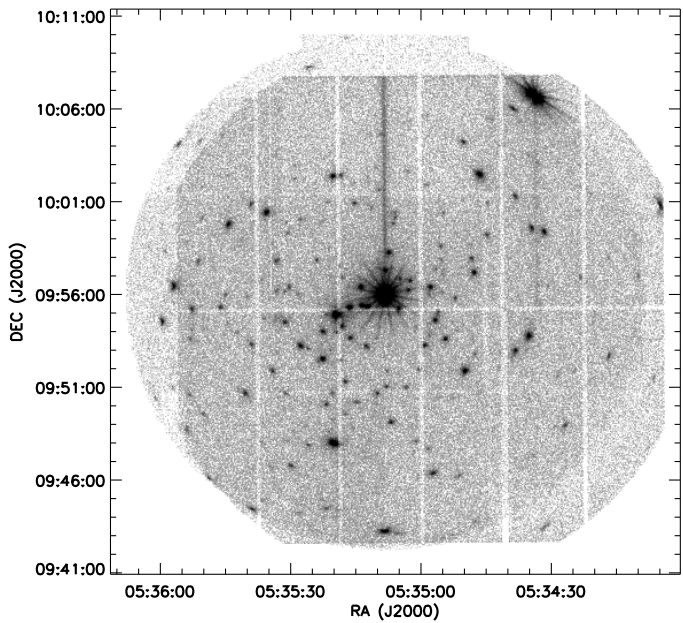}}
\caption{Composite EPIC MOS1+MOS2+PN image of the $\lambda$~Ori field in
the 0.3$-$7.8~keV energy band.}
\label{epic_ima}
\end{figure}

\section{Observations and data analysis}
\label{observ}

The $\lambda$~Ori cluster was observed by {\em XMM-Newton} from 20:46 UT on
September 28, 2006 to 12:23 UT on September 29, 2006 (Obs. ID 0402050101),
for a total duration of 56 ks, using both the EPIC MOS and PN cameras and
the RGS instruments. The EPIC cameras were operated in full frame mode with
the thick filter.

Data analysis was carried out using the standard tasks in SAS v.7.1.0. The
PN data were time-filtered to exclude a period of high background due to
proton flares at the end of the observation. The final effective exposure
time was $\sim$\,55~ks for each MOS and $\sim$\,53~ks for PN. We limited our
analysis to the 0.3$-$7.8~keV energy band to exclude low-energy events, which
are mostly noise and artifacts, and the background, which dominates the
emission above 7.8~keV. The image of the combined MOS1+MOS2+PN events in the
0.3$-$7.8~keV energy band is shown in Fig.~\ref{epic_ima}.

\subsection{Source detection}
\label{detection}

Source detection was performed both on the individual datasets and on the
merged MOS1+MOS2+PN dataset using the wavelet detection algorithm developed
at INAF\,--\,Osservatorio Astronomico di Palermo \citep{damiani97}, adapted
to the EPIC case. The EPIC version was specifically designed to handle in a
straightforward way source detection on the sum of datasets from different
instruments. We used a detection threshold of $5\sigma$, which ensures at
most one spurious detection in each dataset, and which was determined from a
set of 100 Monte-Carlo simulations of pure background datasets with the same
number of counts as the observation. To take the different sensitivities of
the PN and MOS cameras into account, in the detection on the summed dataset
we scaled the PN exposure map by a factor of $3.1$, derived from the median
ratio of PN to MOS count rates of common sources detected on the individual
datasets. Count rates derived from the detection on the summed dataset are
expressed as MOS equivalent count rates.

After removing a few obviously spurious detections (due to hot pixels, to
the point spread function structure of the central bright source, to
out-of-time events, or to sources split by CCD gaps), we obtained a total of
167 sources, three of which were only detected on a single instrument. To
check for systematic offsets in the derived X-ray coordinates, we
cross-correlated the source list with the 2MASS catalogue \citep{2mass},
using an identification radius of 6\,arcsec. We found a median offset
between the X-ray and optical positions of $-1.8$\,arcsec in right ascension
and $-0.2$\,arcsec in declination. This offset was then used to correct the
X-ray coordinates before performing the source identification. The detected
sources with the corrected X-ray coordinates are listed in
Table~\ref{det_all}, where we also indicate their counterparts identified in
Sect.~\ref{ident}.

\subsection{Extraction and analysis of light curves and spectra}
\label{extraction}

For all sources we extracted light curves from the MOS and PN event files,
using circular regions with radii ranging between $24\arcsec$ for the
brightest, isolated sources, down to $10\arcsec$ for very close sources to
avoid mutual contamination. To investigate the source variability, we
applied a Kolmogorov-Smirnov test on the unbinned photon arrival times for
the combined PN and MOS data. To this aim, we selected only events occurring
in the ``good time intervals'' in common between all instruments. For
sources falling on CCD gaps or close to the CCD edges in one of the
instruments, we excluded data from that instrument to avoid possible
spurious effects. The results of the variability analysis are discussed in
Sect.~\ref{var}.

Spectral analysis was performed for sources with at least 500 counts in the
PN, or in the MOS if PN was not available. PN and MOS spectra were extracted
from the same circular regions used for the light curves. We excluded from
the analysis the sources located on the point-spread-function wings of
$\lambda$~Ori\,AB, since their spectra below 1~keV are strongly contaminated
by the emission from the central hot star. In a few cases we also excluded
spectra from either the PN or the MOS cameras for sources located on a CCD
gap or at the CCD edge of the instrument, since the effective number of
source counts from that instrument is considerably reduced. The only
exception was made for source LOX\,1, which falls close to the edge of the
PN detector but is outside the field of view of both MOS cameras. Background
spectra were extracted from nearby circular regions free from other X-ray
sources and on the same CCD chip, using the same extraction radius as the
corresponding source region. Response matrices and ancillary files were
generated for each source using the standard SAS tasks {\sc rmfgen} and {\sc
arfgen}. Spectra were rebinned to a minimum of 20 counts per bin and were
fitted in XSPEC v.12.5.0. For each source, we performed joint fits of the
available PN and MOS spectra using the APEC v.1.3.0 thermal plasma model
with one or more temperature components, and the WABS model to account for
interstellar absorption. Abundances were left free to vary, and values are
relative to the solar abundances by \cite{anders89}. Errors for each
parameter were computed for $\Delta\chi^2=2.706$.

To investigate the nature of fainter sources, we also computed hardness
ratios for all sources, using background-subtracted counts extracted from
the same circular regions defined above. Counts were extracted from the PN
dataset (or MOS for sources outside the PN field of view or on CCD gaps) in
the following energy bands: 0.3$-$1.0~keV (soft, $S$), 1.0$-$2.4~keV
(medium, $M$) and 2.4$-$7.8~keV (hard, $H$). Hardness ratios were then
defined as HR$_1 = (M-S)/(M+S)$ and HR$_2=(H-M)/(H+S)$. This choice allows
us to distinguish between stellar sources, which are generally soft and emit
most of their luminosity below $\sim$\,1~keV, and highly-absorbed
extragalactic sources, whose emission would mainly be in the higher-energy
bands.

\subsection{Optical catalogue}
\label{opt_cat}

To identify the detected sources, we constructed an optical catalogue of
known objects in the {\em XMM-Newton} field of view from the literature.
\citet{dm99,dm01,dm02} performed an extensive photometric and
medium-resolution spectroscopic survey of the entire SFR, finding a total of
266 late-type members, 72 of which were located in the central 1\,deg region
around $\lambda$~Ori. \citet{barrado04,barrado07}, using deep
optical/infrared photometry, low-resolution spectroscopy, and {\em Spitzer}
imaging, extended the known cluster population to very-low mass stars and
brown dwarfs, finding $\sim$\,150 members and photometric candidates down to
$\sim\,0.02\,M_\odot$. High-resolution spectroscopy by \citet{sacco08} and
\citet{maxted08} provided accurate membership information for $\sim$\,90
low-mass candidates. Recently, \citet{bouy09lam} performed a deep
near-infrared survey of the central 5 arcmin of the cluster, finding nine
new very low-mass member candidates and a faint visual companion to
$\lambda$~Ori\,C. We added additional bright stars from the photometric
study by \citet{murdin77} and from the X-ray study by \citet{stone85}.
Proper motion membership for bright stars is provided by \citet{dias01} and
\citet{kharchenko04}. Most of the stars in the catalogue, with the exception
of a few faint objects, have 2MASS counterparts, so we used their
2MASS coordinates to have more accurate positions.

The final catalogue contains 153 stars falling in the {\em XMM-Newton} field
of view, 128 of which are probable or possible cluster members. 85 late-type
members are spectroscopically-confirmed, showing both the presence of youth
indicators (strong Li~{\sc i} and/or weak Na~{\sc i} absorption lines), as
well as radial velocity consistent with that of the cluster
\citep{dm99,sacco08,maxted08}. Six of the cluster members are early-type
(O-B-A) stars, including the O8\,III and B0\,V components of
$\lambda$~Ori\,AB, and the Herbig~Ae/Be (HAeBe) star HD\,245185.
{\em Spitzer} data are available for five of the early-type members
\citep{hernandez09}, and for 93 of the late-type members and candidates
\citep{barrado07}, among which 33 have optically thick (Class~II) or
evolved, optically thin discs (EV). However, only nine of the stars with
discs in our sample are known to be accreting from spectroscopic
observations.

For all cluster members and candidates we derived masses, ages, and
bolometric luminosities from the available colour-magnitude diagrams, using
the \citet{siess00} evolutionary tracks with the \citet{kh95} colour
transformations for stars brighter than $I_\mathrm{c}\sim$\,16~mag, and the
\citet{baraffe98} tracks for fainter stars. We used primarily the $I$ vs.
$(R-I)$ diagram for brighter stars, and the $I$ vs. $(I-J)$ diagram for
faint stars. For a few very-low mass stars and brown dwarfs falling outside
the model grids, we estimated the masses and luminosities from the $J$ (or
$K_\mathrm{s}$) magnitude assuming an age of 5~Myr; similarly, we used the
$V$ magnitude and the spectral type to estimate the mass and luminosity of
$\lambda$~Ori\,C for the same age. For $\lambda$~Ori\,AB, the mass and
luminosity were taken from \cite{dm01}. 

\begin{figure}
\resizebox{\hsize}{!}{\includegraphics[clip]{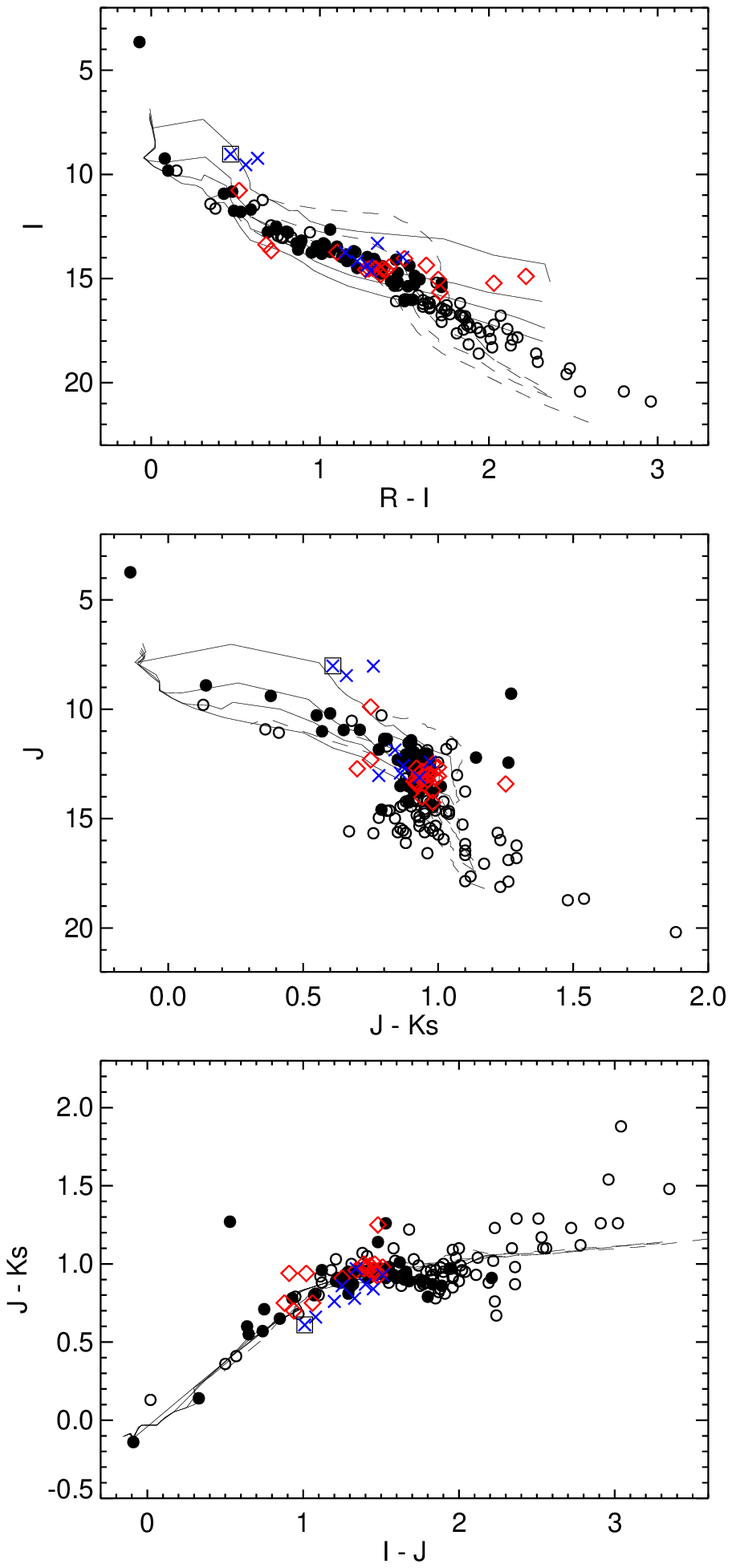}}
\caption{$I$ vs. $(R-I)$ colour-magnitude diagram ({\em top panel}), $J$ vs.
$(J-K_\mathrm{s})$ colour-magnitude diagram ({\em middle panel}), and
$(J-K_\mathrm{s})$ vs. $(I-J)$ colour-colour diagram ({\em bottom panel})
for objects in the {\em XMM-Newton} field of view. Filled and open circles
indicate detected and undetected members and candidates, respectively, while
red diamonds indicate the possible new candidates detected in our
observation. Detected non-members are marked as blue crosses; the boxed
cross symbol indicates HD\,245059, discussed in Sect.~\ref{spec_hd}. The
solid and dashed lines are 1, 5, 10, and 20~Myr isochrones from
\citet{siess00} and \citet{baraffe98}. The star with strong
$(J-K_\mathrm{s})$ excess is the HAeBe star HD\,245185.}
\label{cmd}
\end{figure}


\begin{table*}[!ht]
\tiny
\centering
\caption{\label{det_mem}
X-ray and optical properties of sources identified with known cluster
members and candidates.}
\begin{tabular}{rllcrrrrrrrllc}
\hline\hline
LOX & Identification\tablefootmark{a}& $\log L_\mathrm{X}$\tablefootmark{b}& 
$\log L_\mathrm{X}/L_\mathrm{bol}$& $V$& $R$& $I$& $J$& $H$& $K_\mathrm{s}$&
Mass & Sp.T& Disc& Sp\tablefootmark{d}\\
& & (erg/s) & & (mag)& (mag)& (mag)& (mag)& (mag)& (mag)& ($M_\odot$)& & class\tablefootmark{c}& \\
\hline
  1& 2M J05340691+1001005   & 30.52*& $-3.70$& 11.82 & 11.31 & 10.83 & 10.19&  9.74&  9.59&  1.72& \ldots& -    & -\\
 12& DM 9                   & 30.15*& $-3.26$& 14.21 & 13.46 & 12.77 & 11.84& 11.23& 11.06&  1.10& \ldots& -    & s\\
 16& DM 11                  & 29.89 & $-3.08$& 16.42 & 15.17 & 13.99 & 12.46& 11.73& 11.49&  0.42& \ldots& III  & s\\
 17& DM 12                  & 30.40*& $-2.77$& 15.91 & 14.57 & 13.47 & 12.05& 11.32& 11.09&  0.48& \ldots& III  & s\\
 19& DM 14                  & 30.02*& $-3.16$& 15.09 & 14.17 & 13.31 & 12.07& 11.36& 11.19&  0.78& \ldots& -    & s\\
 21& DM 16                  & 30.03 & $-2.99$& 16.08 & 15.02 & 13.86 & 12.37& 11.66& 11.43&  0.44& \ldots& -    & s\\
 26& DM 18                  & 30.57*& $-2.94$& 14.01 & 13.25 & 12.51 & 11.44& 10.80& 10.64&  1.06& \ldots& -    & s\\
 27& LOri 068               & 29.47 & $-3.07$& 19.28 & 16.76 & 15.20 & 13.52& 12.90& 12.63&  0.29& M5.0  & III  & s\\
 29& DM 19                  & 29.91*& $-3.16$& 15.79 & 14.78 & 13.72 & 12.42& 11.72& 11.54&  0.50& \ldots& -    & s\\
 31& DM 22                  & 30.26*& $-2.78$& 15.58 & 14.70 & 13.75 & 12.43& 11.77& 11.56&  0.63& \ldots& -    & s\\
 37& LOri 075               & 28.69 & $-3.87$& \ldots& 16.95 & 15.23 & 13.40& 12.79& 12.53&  0.26& M5.0  & III  & -\\
 38& LOri-SOC-1             & 28.87 & $-3.14$& \ldots& \ldots& 16.39 & 14.59& 14.16& 13.80&  0.20& \ldots& -    & -\\
 41& LOri 050               & 29.29 & $-3.49$& \ldots& 15.90 & 14.54 & 12.88& 12.24& 11.95&  0.36& M4.5  & II a & s\\
 42& LOri-SOC-2             & 28.93 & $-3.30$& \ldots& \ldots& 16.21 & 14.27& 13.67& 13.30&  0.16& \ldots& -    & -\\
 44& LOri 024               & 29.85*& $-3.32$& \ldots& 14.43 & 13.45 & 12.14& 11.45& 11.22&  0.58& \ldots& III  & s\\
 46& HD 245140              & 29.91*& $-5.34$&  9.22 &  9.32 &  9.24 &  8.91&  8.83&  8.77&  2.60& B9    & no disc& -\\
 47& LOri 056               & 28.96 & $-3.71$& \ldots& 16.43 & 14.87 & 13.21& 12.57& 12.27&  0.30& M4.5  & III  & s\\
 55& LOri 043               & 29.46 & $-3.46$& \ldots& 15.46 & 14.16 & 12.71& 12.02& 11.74&  0.38& \ldots& III  & s\\
 62& DM 24                  & 29.01 & $-4.84$& 12.77 & 12.24 & 11.75 & 11.01& 10.53& 10.44&  1.30& \ldots& -    & s\\
 65& LOri 066               & 28.78 & $-3.71$& \ldots& 17.12 & 15.40 & 13.51& 12.90& 12.65&  0.26& \ldots& III  & s\\
 66& DM 25                  & 29.78 & $-3.38$& 15.40 & 14.26 & 13.38 & 12.26& 11.56& 11.30&  0.74& \ldots& III  & s\\
 67& LOri 045               & 29.94*& $-2.95$& \ldots& 15.56 & 14.23 & 12.77& 12.10& 11.84&  0.38& \ldots& III  & s\\
 70& $\lambda$ Ori C        & 29.98 & $-4.42$& 11.2  & \ldots& \ldots&  9.39&  9.11&  9.01&  2.0 & F8   & -    & -\\
 71& $\lambda$ Ori AB       & 32.31*& $-6.73$&  3.53 &  3.58 &  3.65 &  3.74&  3.77&  3.88& 26.8 & O8III & no disk& -\\
 73& DM 26                  & 30.56*& $-3.26$& 12.92 & 12.33 & 11.80 & 10.95& 10.43& 10.30&  1.42& \ldots& -    & s\\
 75& HD 245185              & 28.81 & $-6.21$&  9.96 &  9.92 &  9.82 &  9.29&  8.76&  8.02&  2.20& A5    & HAeBe& -\\
 78& LOri 057               & 29.06 & $-3.55$& \ldots& 16.63 & 15.04 & 13.41& 12.77& 12.49&  0.29& M5.5  & III  & s\\
 80& LOri 048               & 29.55 & $-3.27$& \ldots& 15.78 & 14.41 & 12.89& 12.20& 11.93&  0.36& \ldots& EV n & s\\
 81& LOri 016               & 29.92 & $-3.33$& \ldots& 14.07 & 13.18 & 11.96& 11.28& 11.05&  0.71& \ldots& III  & s\\
 82& LOri 019               & 30.07 & $-3.16$& \ldots& 14.33 & 13.31 & 12.02& 11.32& 11.07&  0.54& \ldots& III  & -\\
 84& DM 29                  & 29.31 & $-3.68$& 16.76 & 15.25 & 13.97 & 12.55& 11.84& 11.61&  0.39& \ldots& III  & s\\
 85& LOri 062               & 28.78 & $-3.75$& 17.94 & 16.62 & 15.16 & 13.63& 13.00& 12.72&  0.31& \ldots& II n & s\\
 87& LOri 006               & 30.38*& $-3.03$& \ldots& 13.55 & 12.75 & 11.54& 10.86& 10.65&  0.90& \ldots& III  & -\\
 89& DM 30                  & 29.55 & $-3.35$& 16.46 & 15.32 & 14.16 & 12.80& 12.06& 11.83&  0.44& \ldots& -    & s\\
 91& LOri 065               & 28.99 & $-3.47$& \ldots& 16.89 & 15.37 & 13.82& 13.12& 12.84&  0.29& \ldots& III  & s\\
 93& LOri 061               & 29.06 & $-3.47$& \ldots& 16.58 & 15.15 & 13.53& 12.83& 12.52&  0.32& \ldots& II a & s\\
 94& DM 32                  & 29.49 & $-3.29$& 16.73 & 15.71 & 14.49 & 13.07& 12.42& 12.16&  0.39& \ldots& -    & s\\
 99& 2M J05351974+0947476   & 30.82*& $-3.37$& 11.85 & 11.36 & 10.93 & 10.28&  9.90&  9.73&  1.40& \ldots& -    & -\\
100& DM 33                  & 30.54*& $-2.83$& 16.18 & 15.10 & 13.97 & 12.44& 11.64& 11.18&  0.46& \ldots& II a & s\\
102& LOri 060               & 28.95 & $-3.59$& 18.21 & 16.56 & 15.14 & 13.60& 12.96& 12.66&  0.32& M4.5  & III  & s\\
106& LOri 055               & 29.59 & $-3.09$& \ldots& 16.12 & 14.76 & 13.18& 12.48& 12.25&  0.36& M4.5  & III  & s\\
108& DM 34                  & 29.39 & $-3.44$& 16.80 & 15.68 & 14.38 & 12.89& 12.14& 11.92&  0.38& \ldots& -    & s\\
110& DM 35                  & 30.08*& $-2.99$& 15.43 & 14.48 & 13.61 & 12.31& 11.63& 11.46&  0.77& \ldots& -    & s\\
114& DM 36                  & 30.34 & $-2.75$& 16.21 & 14.89 & 13.69 & 12.21& 11.46& 11.07&  0.42& \ldots& II a & s\\
125& LOri 080               & 29.06 & $-3.14$& \ldots& 17.51 & 16.01 & 13.80& 13.20& 12.89&  0.28& M5.5  & EV a & s\\
128& DM 38                  & 29.50 & $-3.46$& 17.22 & 15.55 & 14.10 & 12.50& 11.86& 11.59&  0.35& \ldots& III  & s\\
135& DM 39                  & 30.24*& $-2.79$& 15.87 & 14.82 & 13.81 & 12.45& 11.80& 11.50&  0.55& \ldots& III  & s\\
137& DM 40                  & 29.18 & $-3.54$& 17.23 & 16.01 & 14.66 & 13.23& 12.52& 12.27&  0.36& \ldots& -    & s\\
144& DM 41                  & 29.81 & $-3.15$& 16.76 & 15.38 & 14.06 & 12.55& 11.88& 11.59&  0.38& \ldots& III  & s\\
147& LOri 083               & 28.88 & $-3.33$& \ldots& 17.56 & 16.02 & 14.26& 13.64& 13.37&  0.24& \ldots& III  & s\\
151& LOri 004               & 29.68 & $-3.82$& 15.04 & 13.71 & 12.65 & 11.36& 10.78& 10.55&  0.50& \ldots& III  & -\\
156& DM 44                  & 30.06 & $-3.14$& 15.69 & 14.41 & 13.38 & 12.10& 11.41& 11.16&  0.53& \ldots& III  & s\\
159& LOri 064               & 29.41 & $-3.05$& 18.15 & 16.78 & 15.34 & 13.78& 13.10& 12.85&  0.30& \ldots& EV   & s\\
160& LOri 054               & 29.71 & $-3.00$& 17.86 & 16.19 & 14.73 & 13.19& 12.51& 12.27&  0.33& \ldots& III  & s\\
161& DM 45                  & 30.26 & $-3.60$& 12.89 & 12.28 & 11.69 & 10.94& 10.38& 10.23&  1.48& \ldots& -    & s\\
162& DM 46                  & 30.63*& $-2.82$& 14.43 & 13.39 & 12.65 & 11.42& 10.72& 10.52&  1.05& \ldots& III  & s\\
163& DM 47                  & 29.48 & $-3.38$& 17.41 & 15.91 & 14.38 & 12.73& 12.10& 11.83&  0.33& \ldots& III  & s\\
166& DM 51                  & 30.45 & $-2.94$& 14.86 & 13.60 & 12.79 & 11.55& 10.86& 10.65&  0.88& \ldots& III  & s\\
\hline
\end{tabular}
\tablefoot{
\tablefoottext{a} Identifications labeled DM, LOri and LOri-SOC are from
\citet{dm99}, \citet{barrado04} and \citet{bouy09lam}, respectively. The
2MASS objects (2M) are stars X2 and X4 from \citet{stone85}.
\tablefoottext{b} X-ray luminosity in the 0.3$-$8~keV band; values marked
with an asterisk are derived from spectral fits.
\tablefoottext{c} Disk classification from \citet{barrado07} and
\citet{hernandez09}; for stars with discs (II or EV) we also indicate
whether there is spectroscopic evidence for accretion (a) or non accretion (n).
\tablefoottext{d} An ``s'' in this column indicates
spectroscopically-confirmed members (i.e. objects with spectroscopic youth
features and radial velocity consistent with membership).
}
\end{table*}


\subsection{Source identification}
\label{ident}

We cross-correlated the X-ray source list with the optical catalogue using a
search radius of 4~arcsec. This radius was determined by constructing the
cumulative distribution of the offsets between X-ray and optical position,
following \citet{randsch95}; with this value, we expected to have at most
three spurious identifications. We found 67 sources with at least one
optical counterpart in our catalogue, 58 of which were identified with
cluster members or candidates (indicated with ``Member'' in
Table~\ref{det_all}). Three sources were identified with early-type members
with $M>2\,M_\odot$: $\lambda$~Ori\,AB (the two components are too close to
be resolved by {\em XMM-Newton}), the B9 star \object{HD\,245140}, and the
HAeBe star \object{HD\,245185}. We detected $\sim$\,60\% (9/15) of the stars
with $M=1.0-2.0\,M_\odot$, and $\sim$\,74\% of stars between 0.25 and
$1.0\,M_\odot$ (43/58). Only three of the 58 objects with $M<0.25\,M_\odot$
were detected, \object{LOri\,083}, \object{LOri-SOC-1}, and
\object{LOri-SOC-2}, with masses of 0.16$-$0.24\,$M_\odot$ and X-ray
luminosities of $\sim\,8\times 10^{28}$~erg~s$^{-1}$. The X-ray and optical
properties of the sources identified with cluster members and candidates are
listed in Table~\ref{det_mem}. 

One of the sources, LOX\,70, identified with the F8V star
\object{$\lambda$~Ori\,C}, has another faint counterpart within the
identification radius, \object{LOri-MAD-30}. This object was recently
discovered by \citet{bouy09lam} as a close, very-low mass visual companion
to $\lambda$~Ori\,C. Assuming it belongs to the cluster, they estimate a
mass of $\sim$\,0.04\,$M_\odot$ for an age of 5~Myr from its magnitude
($K_\mathrm{s} = 14.75$~mag), i.e. it would be a candidate brown dwarf. The
X-ray emission from such a very-low mass object is expected to fall below
the detection limit (see Sect.~\ref{xlum} and Fig.~\ref{lx-mass}), and it
would therefore be more than one order of magnitude lower than the observed
count rate. LOX\,70 underwent a flare during the observation with an
increase in the count rate by more than a factor of 2 (see Sect.~\ref{var}).
Unfortunately, the strong contamination from $\lambda$~Ori\,AB hides the
quiescent emission level of the source, preventing a definitive conclusion
about the origin of the X-ray emission. However, since the luminosity of
LOX\,70 is consistent with that of other stars of masses similar to
$\lambda$~Ori\,C ($\sim$\,2\,$M_\odot$), we believe it more likely that the
observed emission is associated with the F8 star. Therefore, we have
assigned all the observed flux to $\lambda$~Ori\,C.

For the remaining 71 undetected cluster members and candidates we computed
$3\sigma$ upper limits at the optical positions using the wavelet algorithm;
their optical and X-ray properties are given in Table~\ref{upplim}. The
location of detected and undetected cluster members in the $I$ vs. $(R-I)$
and $J$ vs. $(J-K_\mathrm{s})$ colour-magnitude diagrams, and in the
$(J-K_\mathrm{s})$ vs. $(I-J)$ colour-colour diagram is shown in
Fig.~\ref{cmd}.

The other nine sources were identified with cluster non-members (indicated
with NM in Table~\ref{det_all}). Three of them (LOX\,7 = \object{LOri\,046},
LOX\,20 = \object{LOri\,036}, and LOX\,45 = \object{LOri\,020}) show no
evidence of strong Li~{\sc i} absorption and are likely field stars
\citep{sacco08}. Additional four stars (LOX\,2 = \object{LOri\,044}, LOX\,3
= \object{LOri\,052}, LOX\,14 = \object{HD\,245059}, and LOX\,30 =
\object{DM\,20}) show signatures of youth (strong Li~{\sc i} absorption
and/or low-gravity lines) but have radial velocity inconsistent with
membership \citep{alcala00,dm99,maxted08}; HD\,245059 also has a proper
motion inconsistent with membership according to \citet{dias01} and
\citet{kharchenko04} and is located significantly above the cluster sequence
in colour-magnitude diagrams. The remaining two objects (LOX\,64 =
\object{TYC\,705-860-1} and LOX\,113 = \object{TYC\,705-937-1}) are located
significantly above the cluster sequence, and we consider them as probable
photometric non-members.


\begin{table*}
\caption{\label{det_cand}
X-ray sources identified with possible new cluster candidates.}
\centering
\begin{tabular}{rclllrrrlcc}
\hline\hline
LOX & 2MASS & \multicolumn{1}{c}{$V$\tablefootmark{a}}&
\multicolumn{1}{c}{$R$\tablefootmark{a}}&
\multicolumn{1}{c}{$I$\tablefootmark{a}}& 
\multicolumn{1}{c}{$J$}& \multicolumn{1}{c}{$H$}&
\multicolumn{1}{c}{$K_\mathrm{s}$}& 
$\log L_\mathrm{X}$\tablefootmark{b}& 
$\log L_\mathrm{X}/L_\mathrm{bol}$\tablefootmark{c}& Mass\tablefootmark{c} \\
& & (mag)& (mag)& (mag)& (mag)& (mag)& (mag)& (erg/s)& & ($M_\odot$) \\
\hline
  4& J05341833+0952376& 15.09 & 14.37 & 13.66 & 12.72 & 12.21& 12.02& 29.80 & $-3.25$& 0.88\\
  8& J05342809+0948476& 17.42 & 16.16 & 14.80 & 13.32 & 12.59& 12.40& 29.78 & $-2.89$& 0.35\\
 34& J05345260+0955500& \ldots& 17.11*& 14.89*& 13.41 & 12.56& 12.16& 29.23 & $-3.41$& 0.33\\
 39& J05345564+0957581& \ldots& 17.25*& 15.22*& 13.71 & 12.96& 12.73& 28.93 & $-3.72$& 0.31\\
 52& J05350064+0951510& \ldots& 14.84*& 13.74*& 12.83 & 12.11& 11.89& 29.28 & $-3.79$& 0.48\\
 56& J05350309+0956162& \ldots& \ldots& \ldots& 12.69 & 12.04& 11.77& 29.65 & $-3.35$& 0.52\\  
 58& J05350356+0950531& \ldots& 16.74*& 15.04*& 14.02 & 13.43& 13.08& 29.36 & $-3.27$& 0.27\\
 59& J05350496+0956561& \ldots& \ldots& \ldots& 13.22 & 12.52& 12.28& 29.33 & $-3.39$& 0.33\\  
 60& J05350528+0955149& \ldots& \ldots& \ldots& 12.50 & 11.75& 11.52& 30.11 & $-2.97$& 0.59\\  
 69& J05350794+0950545& 17.13 & 15.85 & 14.44 & 12.95 & 12.25& 12.02& 29.27 & $-3.55$& 0.35\\
 76& J05351006+0950328& 16.93 & 15.84 & 14.57 & 13.23 & 12.50& 12.27& 29.62 & $-3.13$& 0.38\\
 79& J05351205+0955218& \ldots& \ldots& \ldots& 13.44 & 12.77& 12.52& 30.10 & $-2.48$& 0.25\\  
 83& J05351456+0950026& 17.00 & 15.85 & 14.52 & 13.06 & 12.32& 12.06& 29.34 & $-3.44$& 0.37\\
 86& J05351606+0953374& \ldots& 15.99*& 14.36*& 12.97 & 12.20& 12.00& 29.69 & $-3.20$& 0.31\\
 92& J05351794+0954167& \ldots& 15.54*& 14.04*& 12.65 & 11.88& 11.65& 29.79 & $-3.20$& 0.34\\
 96& J05351857+0944058& 16.83 & 15.78 & 14.47 & 13.02 & 12.31& 12.06& 29.55 & $-3.24$& 0.38\\
 98& J05351857+0944058& 11.85 & 11.29 & 10.77 &  9.89 &  9.38&  9.14& 31.04*& $-3.20$& 1.98\\
107& J05352135+0954549& \ldots& 15.91*& 14.54*& 13.29 & 12.59& 12.38& 28.81 & $-3.96$& 0.36\\
109& J05352216+0953586& 17.07 & 15.82 & 14.53 & 13.08 & 12.38& 12.14& 29.83*& $-2.94$& 0.38\\
112& J05352320+0952190& \ldots& 17.37*& 15.66*& 14.26 & 13.55& 13.28& 28.48 & $-4.02$& 0.16\\
121& J05352846+1002275& 17.18 & 15.97 & 14.58 & 13.15 & 12.45& 12.16& 28.99 & $-3.77$& 0.36\\
123& J05352920+0946317& 14.78 & 14.05 & 13.37 & 12.31 & 11.69& 11.56& 29.73 & $-3.44$& 0.93\\
150& J05354495+0955190& \ldots& \ldots& \ldots& \ldots& 13.95& 13.70& 29.36\tablefootmark{d}& $-2.67$& 0.12\\  
 " & J05354519+0955203& \ldots& \ldots& \ldots& \ldots& 13.71& 13.44& 29.36\tablefootmark{d}& $-2.78$& 0.14\\  
\hline
\end{tabular}
\tablefoot{
\tablefoottext{a} 
Optical photometry from \citet{dm02}, except for the values marked with an
asterisk that are photographic magnitudes from the USNO-B1.0 or GSC2.3
catalogues.
\tablefoottext{b} X-ray luminosity in the $0.3-7.8$ keV band assuming
$d=400$~pc. Values marked by an asterisk are derived from the spectral fits.
\tablefoottext{c} 
For stars with optical photometry, $L_\mathrm{bol}$ and masses were derived
from the optical colours; for the others, they were computed from the $J$ or
$K_\mathrm{s}$ magnitudes for an age of 5~Myr.
\tablefoottext{d} The X-ray luminosity has been equally divided between the
two stars (see text).
}
\end{table*}


We cross-correlated the source list also with the 2MASS and the \citet{dm02}
catalogues, finding counterparts for additional 28 X-ray sources, with two
possible identifications within 4\,arcsec in the case of source LOX~150. All
the counterparts were found in the 2MASS catalogue, and 15 of them have also
optical photometry from \citet{dm02}. Among these objects, 23 have
photometry that is consistent with the cluster sequence in optical and
infrared colour-magnitude diagrams and colour-colour diagrams, as shown in
Fig.~\ref{cmd}, therefore we classify them as possible new candidate
members. We list their properties in Table~\ref{det_cand}, and the other
identifications are indicated in Table~\ref{det_all}.

The two counterparts of LOX~150 are located at 1.1 and 2.8\,arcsec from the
X-ray source. These two objects are very similar with comparable magnitudes
and estimated masses of $\sim$\,0.1$-$0.2\,$M_\odot$. From the inspection of
the X-ray image, it is not possible to determine whether the bulk of the
emission is associated with only one of the two stars or if both contribute
to the source in a comparable way. Therefore, for the following analysis we
divided the X-ray flux equally between them.

\begin{figure}
\resizebox{\hsize}{!}{\includegraphics[clip]{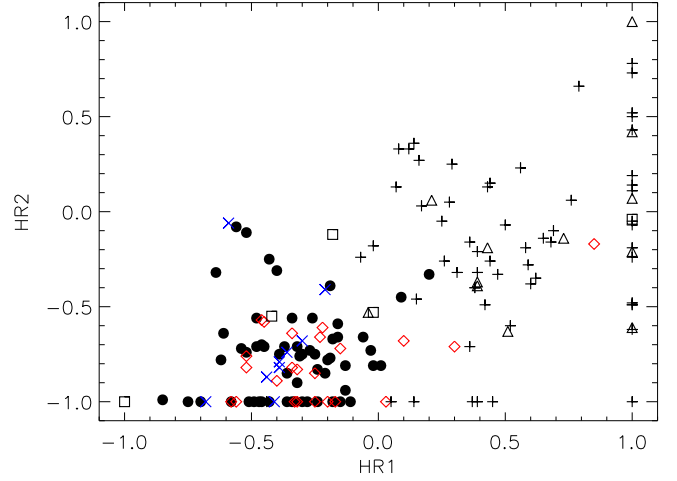}}
\caption{Hardness ratios HR$_2$ vs. HR$_1$ for all detected sources. We
indicate with different symbols cluster members and candidates ({\em filled
circles}), new candidates ({\em red diamonds}), previously known non-members
({\em blue crosses}), other sources with counterparts in 2MASS ({\em open
squares}) or in other catalogues ({\em triangles}) and sources without
counterparts ({\em pluses}).}
\label{hardness}
\end{figure}

To increase the number of identifications, we also searched all available
catalogues in the {\em VizieR} database%
\footnote{Available at {\tt http://vizier.u-strasbg.fr/}}, 
finding counterparts for an additional 12 sources. Five sources have
counterparts in the USNO-B1.0 catalogue, six identifications are found in
the Guide Star Catalogue 2.3 (GSC2.3), all classified as non-stellar
objects, and the remaining one is the radio source 4C~09.21
\citep{day66,williams68}. We list these identifications also in
Table~\ref{det_all}. The remaining 60 sources have no known counterpart in
any astronomical catalogue. It is likely that most of them are background
extragalactic objects. We estimated the expected number of extragalactic
X-ray sources in our observation using the studies by \citet{tozzi01} and
\citet{alex03}. Considering that the sensitivity of our observation ranges
from 0.2~cts~ks$^{-1}$ in the centre of the field to 0.7~cts~ks$^{-1}$ in
the outer regions, and assuming a power-law spectrum with $\Gamma=1.4$ and a
Galactic absorption of $2\times 10^{21}$~cm$^{-2}$ towards $\lambda$~Ori, we
expect $\sim$\,65$-$80 extragalactic X-ray sources in our observation, in
agreement with the number of unidentified sources.

To further check the nature of unidentified sources, in Fig.~\ref{hardness}
we plot the hardness ratios for all detected sources. As expected, all known
cluster members and candidates show soft spectra with HR$_1 \la 0.2$ and
HR$_2 <0$. Similar hardness ratios are found for known cluster non-members
and for all the new candidates except one. On the other hand, most of the
sources with USNO-B1/GSC2.3 identification or without identification have
HR$_1 > 0.2$, indicating harder spectra and supporting their identification
with extragalactic objects.

As mentioned above, one of the new candidates, LOX\,34, shows a
significantly harder emission, with HR$_1 = 0.85$ and HR$_2=-0.17$. This
object is classified as stellar in the GSC2.3 catalogue and is too bright
($R=17.1$~mag, $I=14.9$~mag) to be an extragalactic object. Its position in
the $J$ vs. $(J-K_\mathrm{s})$ and $(J-K_\mathrm{s})$ vs. $(I-J)$ diagrams
indicates the presence of infrared excess. The source is too faint to
perform a reliable spectral analysis ($\sim$\,160~counts in PN); however,
its PN spectrum appears to be consistent with a highly-absorbed coronal
source with a temperature of $\sim$\,1~keV, similar to the values found for
cluster members (see Sect.~\ref{spectra}). The high value of HR$_1$ is a
consequence of the high absorption that strongly reduces the observed
emission below 1~keV. Therefore, we believe that LOX\,34 might be a cluster
member observed through a significant amount of circumstellar material
(maybe an edge-on disk?). Optical spectroscopic observations will be
required to confirm its nature.

\begin{figure}
\resizebox{\hsize}{!}{\includegraphics[clip]{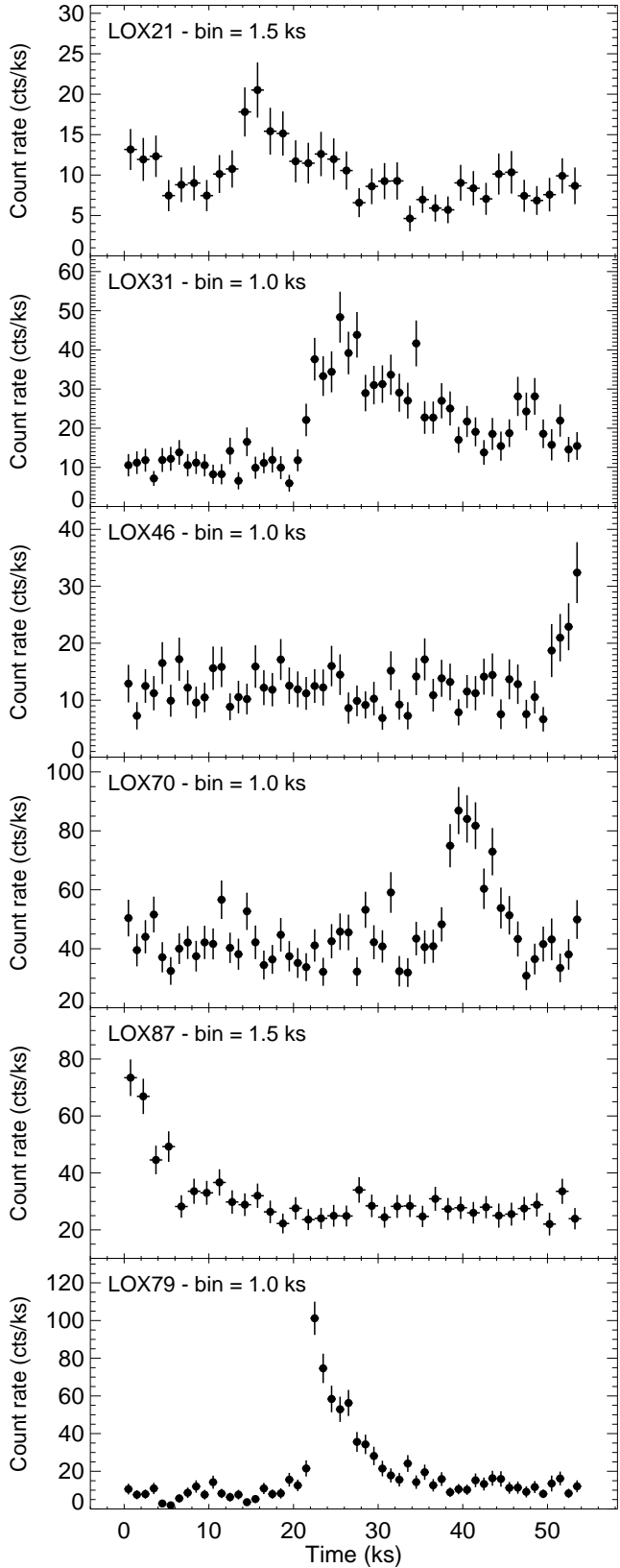}}
\caption{Combined PN+MOS1+MOS2 light curves of cluster members and
candidates showing strong flares during our observation. Count rates are
expressed as MOS equivalent count rates. The different bin size used for
each source is indicated at the top of each panel together with the source
identification.}
\label{lc_flares}
\end{figure}

\begin{figure}
\resizebox{\hsize}{!}{\includegraphics{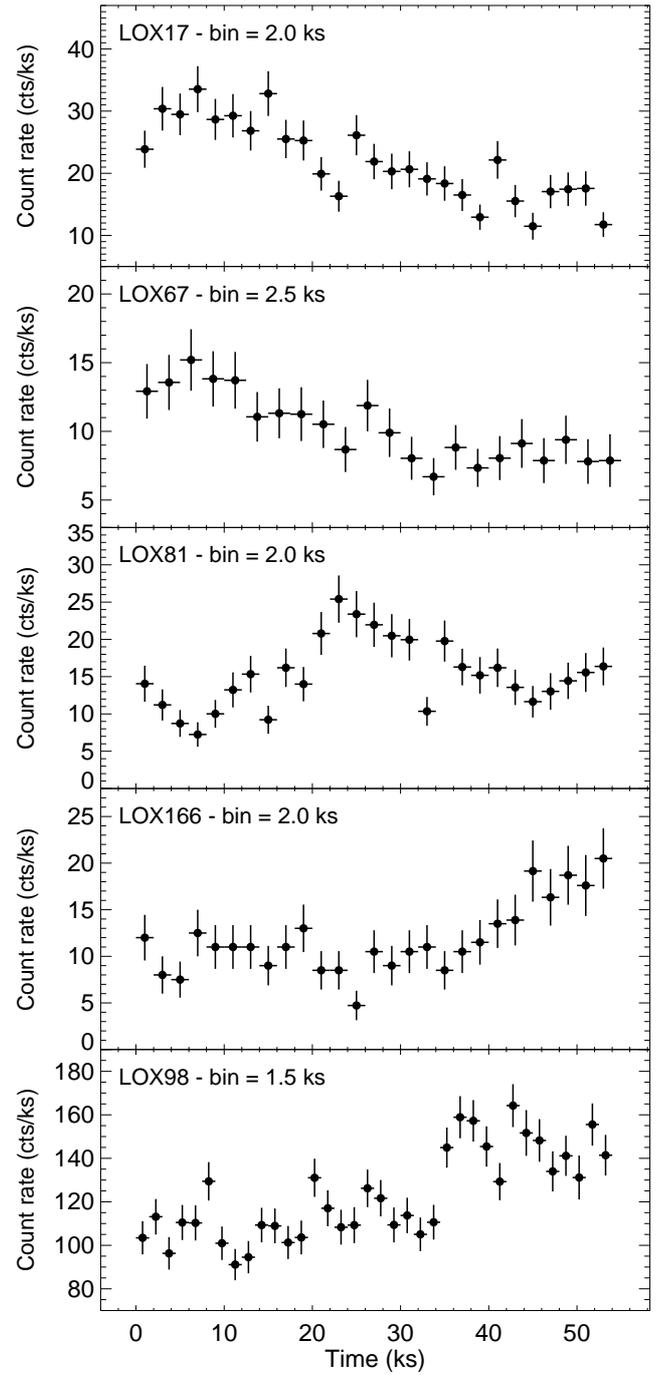}}
\caption{Same as Fig.~\ref{lc_flares} but for the brightest cluster members
and candidates showing gradual variations or variability not clearly
attributable to flares.}
\label{lc_slowvar}
\end{figure}

\section{Variability}
\label{var}

Many of the sources showed significant variability during our observation,
mostly in the form of flares or low-level, irregular variability, as well as
gradual variations over the whole observations. The Kolmogorov-Smirnov test
indicated variability at the 99\% confidence level in 18/58 (31\%) cluster
members and in 8/23 (35\%) new candidates. An additional two cluster members
were variable at the 95\% level. All variable sources are late-type objects,
with the exception of the B9 star LOX\,46 = HD\,245140, which
showed the onset of a strong flare just before the end of the observation.
Significant flares were observed in additional five sources, including one
of the new candidates (LOX\,79). The combined PN+MOS1+MOS2 light curves of
these sources are shown in Fig.~\ref{lc_flares}. These flares show the
typical flare behaviour, commonly observed in active late-type stars, with a
fast rise over $\sim$\,2$-$5~ks followed by a slower decay of
$\sim$\,10$-$20~ks, and increases in the count rate by factors of
$\sim$\,2.5$-$10. 

As mentioned in Sect.~\ref{ident}, in the case of the F8 star LOX\,70 =
$\lambda$~Ori\,C, which lies on the wings of $\lambda$~Ori\,AB, the observed
emission is strongly contaminated by the hot star. Therefore, a significant
fraction (if not all) of the steady emission level observed outside of the
flare is likely due to the contribution from $\lambda$~Ori\,AB. However,
since the emission from the hot star is steady throughout the entire
observation, the flare can be entirely attributed to LOX\,70. Unfortunately,
the fact that the intrinsic quiescent level of LOX\,70 is not known does not
allow us to determine the true strength and duration of the flare.

In Fig.~\ref{lc_slowvar} we plot the light curves of sources showing gradual
variations in the X-ray emission level that are not clearly attributable to
flares. Two of them (LOX\,17 and LOX\,67) show a maximum at the beginning of
the observation followed by decay by a factor of $\sim$\,2, which could
either represent the decay phase of a moderate flare or modulation of the
emission due to stellar rotation. Another source (LOX\,166) shows a gradual
rise by a factor of 2 in $\sim$\,15~ks, while LOX\,81 shows an increase in
the emission level by a factor of $\sim$\,3 with comparable rise and decay
times of 15$-$20~ks. Finally, LOX\,98, identified with a new candidate, shows
a sharp increase by a factor of $\sim$\,1.5 followed by a nearly steady
median emission level. Similar trends are commonly observed in PMS stars
\citep[FPS06;][]{preibisch02,skinner03,favata05coup,ozawa05,francio07xest,getman08coup,caballero10cha},
and can be interpreted as modulation of the emission from active or flaring
regions unevenly distributed on the stellar surface and rotating in and out
of view \citep{stelzer99}.

\section{Spectral analysis}
\label{spectra}

\begin{figure}
\resizebox{\hsize}{!}{\includegraphics{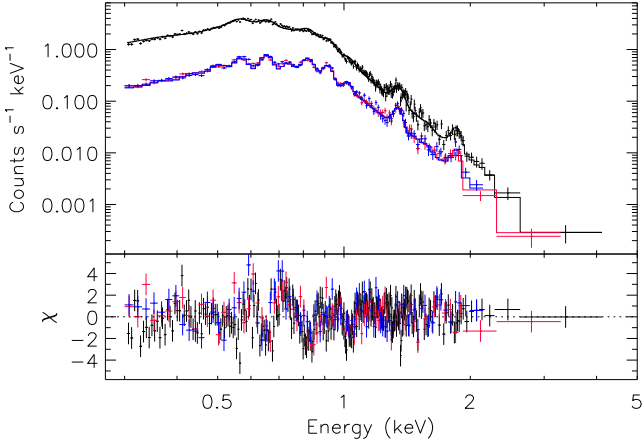}}
\caption{PN (black) and MOS (red and blue) spectra of $\lambda$~Ori\,AB,
together with the best-fit model.}
\label{lori_sp}
\end{figure}

\subsection{$\lambda$~Ori\,AB}
\label{spec_lambda}

The hot star $\lambda$~Ori\,AB (LOX\,71) is the strongest source in the
centre of the field of view. Its PN and MOS spectra are shown in
Fig.~\ref{lori_sp}. The spectrum is soft, with the emission falling below
the background level above $\sim$\,4~keV. We fitted the spectra using three
temperature components and variable individual abundances, and the hydrogen
column density $N_\mathrm{H}$ was left free to vary. The third component was
required to fit the higher energy part of the spectrum, which was slightly
underestimated above 2~keV using only two temperatures.

The resulting best-fit parameters are given in Table~\ref{lori+hd_fit}. The
column density is consistent with the value $N_\mathrm{H}=7.4\times
10^{20}$~cm$^{-2}$ derived from the cluster reddening $E(B-V)=0.12$~mag
\citep{diplas94} using the standard relation $N_\mathrm{H}/A_\mathrm{V}= 2
\times 10^{21}$~cm$^{-2}$~mag$^{-1}$ \citep[and references
therein]{vuong03}. The bulk of the emission is concentrated at temperatures
of $\sim$\,0.2$-$0.3~keV (2$-$3~MK), with equal emission measures. A much
weaker component is present at $\sim$\,0.7~keV (8~MK). Abundances are
subsolar, ranging between 0.3 and 0.5 of the solar values. These results are
consistent with those found for the O9.5V star $\sigma$~Ori\,AB, although
the third component at 0.7~keV was not present in this star
\citep{sanz04sig,skinner08}, and for other hot stars
\citep[e.g.][]{zhekov07,cohen08,naze09}. The unabsorbed X-ray luminosity of
$\lambda$~Ori\,AB in the 0.3$-$8.0~keV band is $L_\mathrm{X}=1.2\times
10^{32}$~erg~s$^{-1}$. 

The star $\lambda$~Ori\,AB was previously observed with {\em ASCA} by
\citet{corcoran94}, who found a significantly hotter spectrum, with a
component at $\sim$\,2.3~keV ($\sim$\,25~MK) in addition to the cooler
emission at 0.3~keV (3~MK), and an iron abundance of $\sim$\,0.2 solar. We
do not find any evidence of such a hot plasma in our observation. However,
the {\em ASCA} spectrum was extracted from a region of $4\arcmin$ radius: as
shown in Fig.~\ref{epic_ima}, there are several bright sources around
$\lambda$~Ori\,AB falling within this radius. If we extract a combined
spectrum of the central hot star and all the other sources included in the
{\em ASCA} extraction region, a hot tail appears, requiring a plasma
component at $\sim$\,2.5~keV (30~MK). We therefore conclude that the hot
plasma inferred from the {\em ASCA} data was not a property of the hot star
itself, but the result of contamination of its spectrum by the unresolved
hotter sources close to it.

\begin{table}
\caption{\label{lori+hd_fit} Result of the joint PN and MOS spectral fitting
for the hot star $\lambda$~Ori\,AB and for the non-member weak-lined T~Tauri
star HD\,245059.}
\centering
\renewcommand{\arraystretch}{1.2}
\begin{tabular}{lll}
\hline\hline
                                    & $\lambda$~Ori\,AB      & HD\,245059\tablefootmark{a}\\
\hline
$N_\mathrm{H}$ ($10^{20}$~cm$^{-2}$)& $6.46_{-0.93}^{+1.12}$ & $1.91_{-1.40}^{+1.04}$\\
$T_1$ (keV)                         & $0.17_{-0.01}^{+0.01}$ & $0.39_{-0.04}^{+0.03}$\\
$T_2$ (keV)                         & $0.29_{-0.01}^{+0.01}$ & $0.81_{-0.05}^{+0.04}$\\
$T_3$ (keV)                         & $0.69_{-0.07}^{+0.05}$ & $1.91_{-0.13}^{+0.13}$\\
$EM_1$ ($10^{54}$~cm$^{-3}$)        & $7.47_{-0.94}^{+1.13}$ & $2.13_{-0.69}^{+0.63}$\\
$EM_2$ ($10^{54}$~cm$^{-3}$)        & $7.51_{-0.94}^{+0.67}$ & $2.93_{-0.48}^{+0.79}$\\
$EM_3$ ($10^{54}$~cm$^{-3}$)        & $0.37_{-0.08}^{+0.17}$ & $3.01_{-0.27}^{+0.34}$\\
O\tablefootmark{b}                  & $0.34_{-0.02}^{+0.02}$ & $0.32_{-0.03}^{+0.09}$\\
Ne\tablefootmark{b}                 & $0.46_{-0.03}^{+0.03}$ & $0.68_{-0.14}^{+0.20}$\\
Mg\tablefootmark{b}                 & $0.50_{-0.06}^{+0.05}$ & $0.37_{-0.09}^{+0.13}$\\
Si\tablefootmark{b}                 & $0.50_{-0.10}^{+0.09}$ & $0.20_{-0.08}^{+0.08}$\\
S\tablefootmark{b}                  & $0.00_{\ldots}^{+0.21}$& $0.17_{-0.16}^{+0.16}$\\
Fe\tablefootmark{b}                 & $0.50_{-0.03}^{+0.04}$ & $0.24_{-0.02}^{+0.08}$\\
$\chi^2_\mathrm{r}$                 & 1.98                   & 1.36                  \\
d.o.f                               & 440                    & 439                   \\
$F_\mathrm{X}$\tablefootmark{c} (erg~cm$^{-2}$~s$^{-1}$)& 
                                      $6.4 \times 10^{-12}$  & $4.4 \times 10^{-12}$ \\
$L_\mathrm{X}$\tablefootmark{c} (erg~s$^{-1}$)& $1.2 \times 10^{32}$& $8.4 \times 10^{31}$  \\
\hline
\end{tabular}
\tablefoot{
\tablefoottext{a} Emission measures and luminosity for HD\,245059 are
computed for $d=400$~pc and are only indicative. If the star were located at
90~pc (see Sect.~\ref{spec_hd}) these values would be reduced by a factor of
20.
\tablefoottext{b} Abundances relative to the solar abundances by
\citet{anders89}.
\tablefoottext{c} Unabsorbed X-ray flux and luminosity in the 0.3$-$8.0~keV
band.
}
\end{table}


\begin{table*}
\caption{\label{fits} Best-fit parameters from the joint PN and MOS spectral
fits for bright cluster members and two new candidates.} 
\centering
\renewcommand{\arraystretch}{1.2}
\begin{tabular}{rccccccrrrl}
\hline\hline
LOX& $T_1$ & $T_2$ & $EM_1$\tablefootmark{a}& $EM_2$\tablefootmark{a}& $Z/Z_\odot$ & $\chi^2_\mathrm{r}$& d.o.f.&
$F_\mathrm{X}$\tablefootmark{b} & $L_\mathrm{X}$\tablefootmark{c}& Notes\\
   & \multicolumn{2}{c}{(keV)}& \multicolumn{2}{c}{($10^{53}$ cm$^{-3}$)}& &
& & & & \\
\hline
  1& $0.56_{-0.14}^{+0.12}$& $1.12_{-0.17}^{+0.24}$& $1.55_{-0.54}^{+0.98}$& $2.28_{-0.69}^{+1.09}$& $0.22_{-0.08}^{+0.19}$& 1.13&  38& 17.4& 3.3& PN only\\
 12& $0.32_{-0.05}^{+0.07}$& $1.18_{-0.18}^{+0.14}$& $0.96_{-0.34}^{+0.48}$& $1.05_{-0.31}^{+0.42}$& $0.21_{-0.10}^{+0.18}$& 0.66&  41&  7.4& 1.4& \\
 17& $0.75_{-0.13}^{+0.10}$& $1.61_{-0.30}^{+1.12}$& $1.32_{-0.54}^{+1.28}$& $1.84_{-0.88}^{+0.57}$& $0.13_{-0.03}^{+0.09}$& 1.08&  84& 13.1& 2.5& \\
 19& $0.81_{-0.06}^{+0.08}$& \ldots                & $1.78_{-0.33}^{+0.34}$& \ldots                & $0.07_{-0.03}^{+0.04}$& 1.26&  37&  5.4& 1.0& \\
 26& $0.37_{-0.06}^{+0.17}$& $1.12_{-0.09}^{+0.08}$& $1.24_{-0.54}^{+0.40}$& $3.83_{-0.56}^{+0.82}$& $0.16_{-0.05}^{+0.06}$& 1.00& 100& 19.3& 3.7& No MOS1\\
 29& $0.43_{-0.09}^{+0.17}$& $1.03_{-0.11}^{+0.16}$& $0.42_{-0.13}^{+0.17}$& $0.56_{-0.21}^{+0.25}$& $0.24_{-0.09}^{+0.22}$& 1.62&  44&  4.3& 0.8& \\
 31& $0.59_{-0.22}^{+0.19}$& $1.58_{-0.25}^{+0.56}$& $0.86_{-0.27}^{+0.31}$& $1.59_{-0.36}^{+0.33}$& $0.10_{-0.04}^{+0.08}$& 1.09&  75&  9.4& 1.8& Flare\\
 44& $0.59_{-0.22}^{+0.15}$& $1.29_{-0.18}^{+0.33}$& $0.13_{-0.04}^{+0.10}$& $0.36_{-0.17}^{+0.02}$& $0.57_{-0.29}^{+1.02}$& 0.74&  31&  3.7& 0.7& \\
 46& $0.54_{-0.10}^{+0.08}$& $1.00_{-0.25}^{+0.71}$& $0.73_{-0.29}^{+0.44}$& $0.31_{-0.21}^{+0.19}$& $0.20_{-0.08}^{+0.17}$& 1.09&  40&  4.3& 0.8& \\
 67& $0.30_{-0.09}^{+0.18}$& $1.04_{-0.14}^{+0.23}$& $0.63_{-0.31}^{+0.33}$& $0.90_{-0.44}^{+0.38}$& $0.11_{-0.05}^{+0.14}$& 0.93&  45&  4.5& 0.9& \\
 73& $0.80_{-0.04}^{+0.05}$& $1.89_{-0.53}^{+2.43}$& $2.85_{-0.59}^{+0.48}$& $1.55_{-0.59}^{+0.63}$& $0.15_{-0.03}^{+0.05}$& 1.24&  81& 18.9& 3.6& \\
 87& $0.49_{-0.12}^{+0.14}$& $1.26_{-0.17}^{+0.23}$& $1.47_{-0.50}^{+0.67}$& $1.53_{-0.40}^{+0.42}$& $0.21_{-0.08}^{+0.15}$& 1.17&  44& 12.7& 2.4& Flare; MOS1+2 only\\
 99& $0.75_{-0.05}^{+0.04}$& $1.60_{-0.15}^{+0.21}$& $3.18_{-0.67}^{+0.84}$& $4.04_{-0.54}^{+0.57}$& $0.22_{-0.05}^{+0.06}$& 0.97& 201& 34.4& 6.6& \\
100& $0.32_{-0.03}^{+0.08}$& $2.72_{-0.92}^{+4.42}$& $2.09_{-1.00}^{+1.13}$& $0.59_{-0.21}^{+0.17}$& $0.12_{-0.07}^{+0.15}$& 1.29&  37&  7.4& 1.4& CTTS; no MOS1\\ 
110& $0.27_{-0.12}^{+0.15}$& $0.82_{-0.04}^{+0.03}$& $0.40_{-0.31}^{+0.54}$& $1.46_{-0.29}^{+0.38}$& $0.14_{-0.05}^{+0.05}$& 0.80&  58&  6.3& 1.3& \\
135& $0.34_{-0.06}^{+0.12}$& $1.22_{-0.17}^{+0.12}$& $1.07_{-0.31}^{+0.42}$& $1.76_{-0.31}^{+0.44}$& $0.12_{-0.05}^{+0.07}$& 0.91&  69&  9.1& 1.8& \\
162& $0.89_{-0.06}^{+0.07}$& \ldots                & $7.39_{-0.94}^{+1.01}$& \ldots                & $0.05_{-0.01}^{+0.03}$& 1.05&  49& 22.3& 4.3& MOS1+2 only\\
\hline
 98& $0.76_{-0.03}^{+0.03}$& $1.87_{-0.20}^{+0.18}$& $4.50_{-0.94}^{+1.19}$& $6.57_{-0.59}^{+0.61}$& $0.26_{-0.06}^{+0.07}$& 1.27& 172& 57.7& 11.0& MOS1+2 only \\
109& $0.36_{-0.08}^{+0.29}$& $1.18_{-0.23}^{+0.16}$& $0.34_{-0.13}^{+2.22}$& $0.54_{-0.19}^{+0.33}$& $0.21_{-0.12}^{+0.26}$& 0.70&  32&  3.5& 0.7& \\
\hline
\end{tabular}
\tablefoot{
\tablefoottext{a} Emission measures are computed for the cluster distance
also for the two 2MASS candidates.
\tablefoottext{b} Unabsorbed X-ray flux in the 0.3$-$8~keV band, in units of
$10^{-14}$~erg\,cm$^{-2}$\,s$^{-1}$.
\tablefoottext{c} Unabsorbed X-ray luminosity in the 0.3$-$8~keV band, in
units of $10^{30}$~erg\,s$^{-1}$.
}
\end{table*}


\subsection{Other cluster members and candidates}
\label{spec_members}

Apart from $\lambda$~Ori\,AB, the sample selected for spectral analysis
includes 17 cluster members and candidates, all of which are late-type
stars, except for the B9 star LOX\,46 (HD\,245140). Only one of the stars in
the sample, LOX\,100 (\object{DM\,33}), is a Class\,II star with active
accretion. Two of the selected sources, LOX\,31 and 87, showed flares during
the observation; however, their count rates are too low to perform a
time-dependent analysis, so only the spectrum for the entire observation was
fitted. In addition, spectral analysis was also performed for two of the
2MASS candidates (LOX\,98 and 109). 

All the spectra were fitted with a 2-temperature model with variable global
abundance. In two cases, only one temperature component was enough to
describe the spectrum, the other one being unconstrained by the fit. At
first we left the hydrogen column density free to vary; however, we found
that in general it was poorly constrained but consistent with the value
$N_\mathrm{H}=7.4\times 10^{20}$~cm$^{-2}$ for all sources, including the
two new candidates. We therefore repeated the fits keeping $N_\mathrm{H}$
fixed to this value, and the resulting best-fit parameters are given in
Table~\ref{fits}. 

In the case of the classical T~Tauri star LOX\,100, the fit with fixed
$N_\mathrm{H}$ is not good at low energies, although acceptable
($\chi^2_\mathrm{r}=1.3$). However, we were not able to find a reasonable
fit by letting also the column density vary: in fact, we obtained a high
value of $N_\mathrm{H}$ ($7\times 10^{21}$~cm$^{-2}$), a very low abundance
(0.02 solar), and an extremely high emission measure of the cooler component
($EM_1=5\times 10^{56}$~cm$^{-3}$), which is implausible. Fixing the
abundance to $Z=0.1\,Z_\odot$, we obtained
$N_\mathrm{H}=1.2_{-0.5}^{+1.0}\times 10^{21}$~cm$^{-2}$, without any
significant improvement in the fit quality.

All the sources in our sample, including the two new candidates, have
similar coronal properties. We find plasma temperatures of
$T_1\sim\,0.3-0.9$~keV (3$-$10~MK) and $T_2\sim\,0.8-2.7$~keV
(9$-$30~MK), emission measure ratio $EM_2/EM_1\sim\,0.3-3.6$, and
subsolar abundances ($Z\sim\,0.1-0.3\,Z_\odot$), in agreement with those
obtained for the $\sigma$~Ori cluster \citep[FPS06;][]{lopez08} and for
other young clusters and SFRs
\citep[e.g.][]{feigel99,getman05coup,gudel07xest}. The average plasma
temperature, weighted with the emission measure, varies between 0.7 and
1.4~keV (8$-$16~MK). 

The plasma parameters derived for the B9 star LOX\,46 (HD\,245140)
are consistent with those of the other late-type sources. As shown in
Fig.~\ref{lc_flares}, this star underwent a flare at the end of the
observation. Late-B stars are not expected to have either strong winds or
magnetically active coronae, and their X-ray emission is generally
attributed to late-type companions. The presence of flaring activity and the
similarity of the X-ray properties of LOX\,46 to those of later-type stars
support the hypothesis that the observed emission comes from an unknown
companion rather than from the B9 star itself.

\subsection{HD\,245059}
\label{spec_hd}

HD\,245059 (LOX\,14, spectral type G8$-$K3) is the brightest source
in our observation, located at the northwestern edge of the EPIC field of
view (see Fig.~\ref{epic_ima}). This star, first detected in X-rays with
{\em Einstein} \citep{stone85}, has been classified as a weak-lined T~Tauri
star, due to its strong Li~{\sc i} absorption and weak H$\alpha$ emission
\citep{fernandez95,alcala96,alcala00,li98}. However, as mentioned in
Sect.~\ref{ident}, HD\,245059 is located significantly above the cluster
sequence and has a radial velocity of $19.8 \pm 1.0$~km~s$^{-1}$
\citep{alcala00}, inconsistent with that of the $\lambda$~Ori cluster
\citep[$27.0 \pm 0.5$~km~s$^{-1}$;][]{sacco08,maxted08}. Its proper motion
\citep[$\mu_\alpha\cos\delta = 11.5\pm 1.5$~mas~yr$^{-1}$, $\mu_\delta =
-35.3\pm 1.0$~mas~yr$^{-1}$;][]{dias01} is also significantly higher than
that of the cluster \citep[$0.5\pm 2.8$ and $-2.5\pm
2.8$~mas~yr$^{-1}$;][]{dias01}. Recent Keck and {\em Chandra} observations
have resolved it into a visual binary \citep{baldovin09}, although there is
no evidence of radial velocity variations \citep{fekel97,alcala00}. The
radial velocity and proper motion of HD\,245059 appear to be consistent with
those of the recently identified 32~Ori moving group
\citep[$v_\mathrm{r} = 18$~km~s$^{-1}$, $\mu_\alpha\cos\delta =
8$~mas~yr$^{-1}$, $\mu_\delta =-33$~mas~yr$^{-1}$;][]{mamajek07}. This
group, located at a distance of $\sim$\,90~pc, consists of a number of
late-type stars detected in the {\em ROSAT All Sky Survey}
\citep[e.g.][]{alcala00} and comoving with the massive binary 32~Ori; the
estimated age is $\sim$\,25~Myr.

The light curve of HD\,245059 shows a steady increase by a factor of 1.5
during the entire observation, with a small flare at the end. We fitted the
spectra using a three-temperature model with variable individual abundances,
and the best-fit parameters are listed in Table~\ref{lori+hd_fit}.
The plasma temperatures are 0.4, 0.8, and 1.9~keV ($\sim$\,5, 9, and 20~MK)
with comparable emission measures, and abundances are strongly subsolar
(0.2$-$0.4), with the exception of Ne which is higher (0.7). We find a
column density $N_\mathrm{H}\sim\,2\times 10^{20}$~cm$^{-2}$, a factor of
$\sim$\,4 lower than the cluster mean value. Our results are consistent,
within the errors, with those obtained with {\em Chandra} for the combined
spectrum of the two binary components%
\footnote{\citet{baldovin09} find $N_\mathrm{H}= 7.7 \times
10^{19}$~cm$^{-2}$, lower than ours by a factor of 2.5 but consistent within
the errors. The differences in the derived $N_\mathrm{H}$ might be a
consequence of the differences in the fitted abundances, since these
parameters are not independent.}
\citep[Table 3]{baldovin09}. If the star were at the distance of the
$\lambda$~Ori cluster, it would have a combined X-ray luminosity of $8\times
10^{31}$~erg~s$^{-1}$: if we assume the same flux ratio between the two
components found by {\em Chandra}, the brightest one would still have
$L_\mathrm{X}\sim\,6\times 10^{31}$~erg~s$^{-1}$, which is nearly one order
of magnitude higher than cluster members with similar colours. On the other
hand, if HD\,245059 belongs to the 32~Ori group at 90~pc, its combined X-ray
luminosity would decrease to $\sim\,4\times 10^{30}$~erg~s$^{-1}$. According
to the \citet{siess00} models and assuming negligible reddening, as
indicated by the low $N_\mathrm{H}$ derived by us and by \citet{baldovin09},
if located at 90~pc this star would have a mass of $\sim\,1.2\,M_\odot$ and
an age of $\sim$\,15~Myr, in agreement with the estimated age of
$\sim$\,25~Myr for the 32~Ori group \citep{mamajek07}. The derived X-ray
luminosity is also consistent with the values found for solar-type stars in
$\lambda$~Ori. These results support the hypothesis that HD\,245059 is a
foreground young star belonging to the 32~Ori group.

\begin{figure}
\resizebox{\hsize}{!}{\includegraphics[clip]{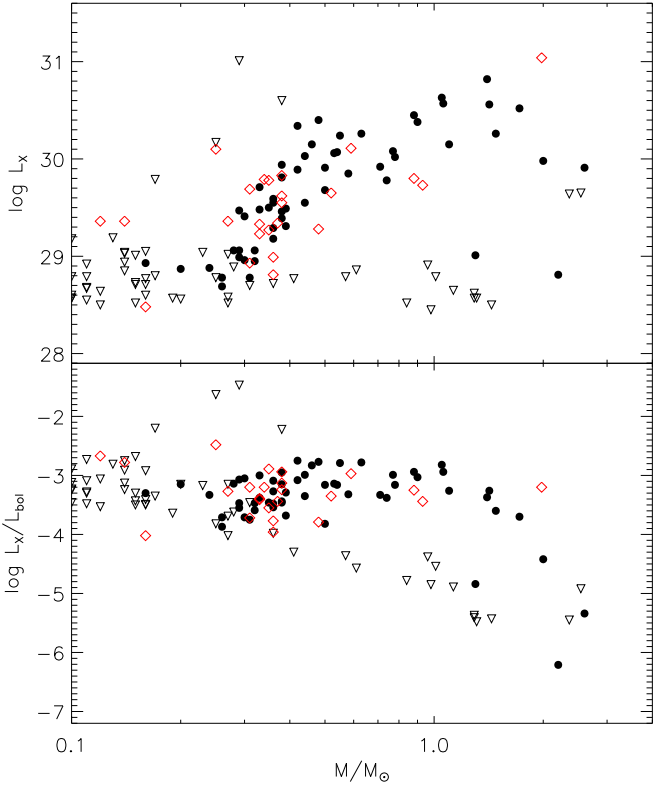}}
\caption{$L_\mathrm{X}$ ({\em top}) and $L_\mathrm{X}/L_\mathrm{bol}$ ({\em
bottom}) as a function of mass for detected (dots) and undetected (open
triangles) cluster members and candidates. We also plot with open diamonds
the new candidates with optical photometry.}
\label{lx-mass}
\end{figure}

\section{X-ray luminosity of cluster members}
\label{xlum}

For the brightest sources for which spectral analysis is available, X-ray
fluxes in the 0.3$-$8.0 keV band were derived directly from the best-fit
models (see Tables~\ref{lori+hd_fit} and \ref{fits}). For the other cluster
members and candidates, we used the results of the best-fit models in
Table~\ref{fits} to derive a count rate to flux conversion factor. To this
aim, we computed the ratio between the unabsorbed X-ray fluxes derived from
the spectral fits and the count rates obtained from the wavelet algorithm on
the summed dataset, and then took the median value. The derived conversion
factor is CF~$=8.7\times 10^{-12}$~erg~cm$^{-2}$~cnt$^{-1}$, with an
uncertainty of $\sim$\,15\% ($1\sigma$ standard deviation). X-ray
luminosities were then computed using the cluster distance of 400~pc.

Applying this conversion factor to the 3\,$\sigma$ upper limits for
undetected members, we obtain a limiting sensitivity $L_\mathrm{X} \sim\,3
\times 10^{28}$~erg~s$^{-1}$ in the centre of the field, decreasing to
$\sim\,6 \times 10^{28}$~erg~s$^{-1}$ at $12\arcmin$ offaxis, and to
$\sim\,1 \times 10^{29}$~erg~s$^{-1}$ in the outer regions covered only by
the MOS cameras.

In Fig.~\ref{lx-mass} we show $\log L_\mathrm{X}$ and $\log
L_\mathrm{X}/L_\mathrm{bol}$ as a function of mass for all cluster members
and candidates, including the new candidates with optical photometry. The
figure does not include the central O star $\lambda$~Ori\,AB
($M\sim\,27\,M_\odot$), which has a luminosity of $\sim
10^{32}$~erg~s$^{-1}$ and $\log L_\mathrm{X}/L_\mathrm{bol} \sim -6.7$,
consistent with the typical value found for hot stars \citep[$\log
L_\mathrm{X}/L_\mathrm{bol} \sim -7$; ][]{pallavic81,bergh97,sana06}. We
find that $L_\mathrm{X}$ increases with stellar mass up to $M \sim
2\,M_\odot$, and then drops at higher masses, as commonly observed in SFRs
and young open clusters. Below $\sim$\,0.3\,$M_\odot$, the X-ray emission of
cluster members drops below the sensitivity limit of our observation,
resulting in the lack of detections at lower masses. We fitted the
relationship below $2\,M_\odot$ using the EM algorithm as implemented in
{\sc asurv} Rev.~1.2 \citep{asurv}, finding $\log L_\mathrm{X} = (1.86 \pm
0.30) \log M/M_\odot + (29.72 \pm 0.14)$, with a standard deviation of
$0.84$. The derived slope is in good agreement with those derived for
Taurus-Auriga \citep[$1.69\pm 0.11$;][]{telleschi07xest} and IC~348
\citep[$1.97 \pm 0.24$;][]{preibisch02}, while it is higher, although
consistent within the errors, than the one found for the Orion Nebula
Cluster \citep[$1.44\pm 0.10$;][]{preibisch05coup}. On the other hand,
$L_\mathrm{X}/L_\mathrm{bol}$ is nearly constant up to $\sim$\,1\,$M_\odot$,
and then decreases at higher masses. The median value for $M\le 1\,M_\odot$
is $\log L_\mathrm{X}/L_\mathrm{bol}\sim -3.3$ for detections, and $\sim
-3.6$ taking upper limits into account, similar to what is found for the
$\sigma$~Ori cluster (FPS06) and for other young clusters and SFRs of
comparable age
\citep[e.g.][]{flaccomio03ev,preibisch05coup,telleschi07xest}. As shown in
Fig.~\ref{lx-mass}, the new candidates fit into these trends very well. Only
one star, \object{DM\,24} = LOX\,62 ($M\sim\,1.4\,M_\odot$), deviates
significantly from the trend, with $L_\mathrm{X}\sim 10^{29}$~erg~s$^{-1}$
and $\log L_\mathrm{X}/L_\mathrm{bol}\sim -5$, one order of magnitude lower
than the other members of similar mass. This star has strong Li~{\sc i}
absorption and variable radial velocity \citep{dm99,dorazi09}, suggesting it
is a spectroscopic binary. Therefore, its membership in the cluster is not
fully confirmed.

\begin{figure}
\resizebox{\hsize}{!}{\includegraphics[clip]{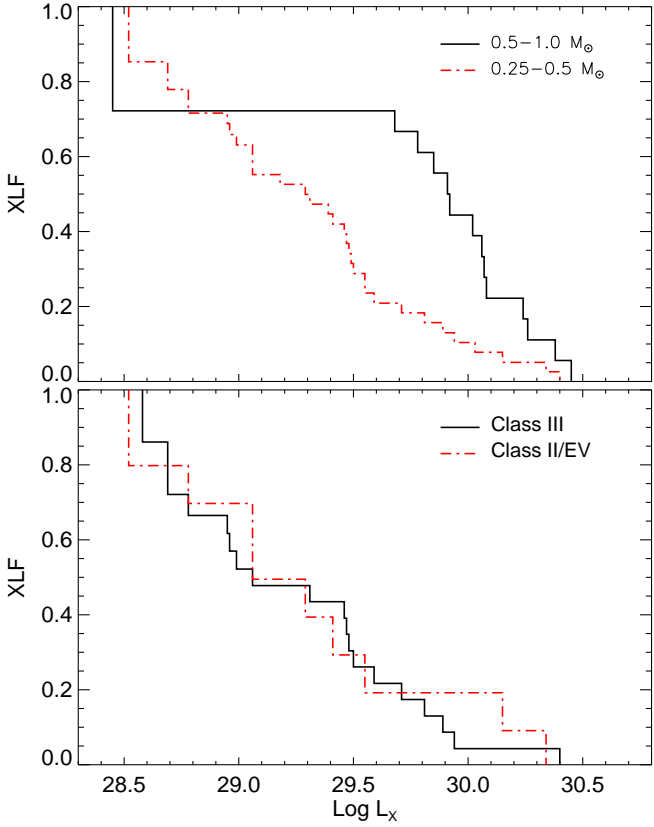}}
\caption{{\em Top panel}: comparison between the XLFs for $\lambda$~Ori
members and candidates with masses in the ranges 0.5$-$1\,$M_\odot$ (black
solid line), and 0.25$-$0.5\,$M_\odot$ (red dash-dotted line).
{\em Bottom panel:} comparison between the XLFs of cluster members and
candidates in the mass range $M=0.25-0.5\,M_\odot$, classified by
\citet{barrado07} as Class~III (black solid line) and Class~II or EV objects
(red dot-dashed line).}
\label{xldf}
\end{figure}

There are 12 stars with masses between 0.4 and 2\,$M_\odot$ that are not
detected, with upper limits of $\sim\,3-8 \times 10^{28}$~erg~s$^{-1}$,
significantly lower than detected members in this mass range. Only one of
them, \object{LOri\,030} ($M\sim\,0.4\,M_\odot$), identified by
\citet{barrado07} as a Class~III source, was observed spectroscopically by
\citet{sacco08}, who classified it as a possible SB2 cluster member. All the
other non-detections are photometric candidates from \citet{murdin77} and
from \citet{barrado04,barrado07}; the latter were all classified as
Class~III sources. Since PMS stars are known to be strong X-ray emitters,
with luminosities $10-10^4$ times higher than older late-type stars
\citep{feigel99}, the lack of strong X-ray emission and of any evidence of
circumstellar material suggests that these objects might be older field
stars unrelated to the $\lambda$~Ori region, rather than true cluster
members.

Figure~\ref{xldf} (top panel) shows the X-ray luminosity functions (XLFs)
for stars in the two mass bins 0.5$-$1\,$M_\odot$ and 0.25$-$0.5\,$M_\odot$,
computed using {\sc asurv}. The difference in X-ray luminosity between the
two samples is clearly evident. We find a median $\log L_\mathrm{X} =
29.9$~erg~s$^{-1}$ for stars between 0.5 and 1\,$M_\odot$, and $\log
L_\mathrm{X} = 29.3$~erg~s$^{-1}$ for stars between 0.25 and 0.5\,$M_\odot$,
i.e. lower by a factor of $\sim$\,4.

In the bottom panel of Fig.~\ref{xldf} we compare the XLF of the sample of
Class~II and EV sources with that of Class~III sources, for stars with
$M=0.25-0.5\,M_\odot$. Higher-mass stars with {\em Spitzer} classification
are all Class~III objects, therefore we exclude them from the comparison to
avoid biases due to their higher luminosities. The two distributions are
indistinguishable, with a median luminosity of $\log L_\mathrm{X} =
29.0$~erg~s$^{-1}$ for both classes, in agreement with the results obtained
for other young clusters and SFRs
\citep[e.g.][]{preibisch01,getman02,feigel03,preibisch05coup}. However, the
sample of stars with discs is very small, containing only 12 objects,
therefore this result might be affected by statistical biases and cannot be
considered conclusive. H$\alpha$ observations are available for 11 Class~II
or EV stars, showing evidence for accretion in six of them
\citep{dm99,sacco08}. Given the small size of the sample of accreting
objects, the comparison of their XLF with that of non-accreting stars is not
useful. Accreting stars were all detected except for one (located too close
to a bright source), with luminosities of $1\times 10^{29} - 2\times
10^{30}$~erg~s$^{-1}$, comparable to those of non-accreting members.

\begin{figure}
\resizebox{\hsize}{!}{\includegraphics[clip]{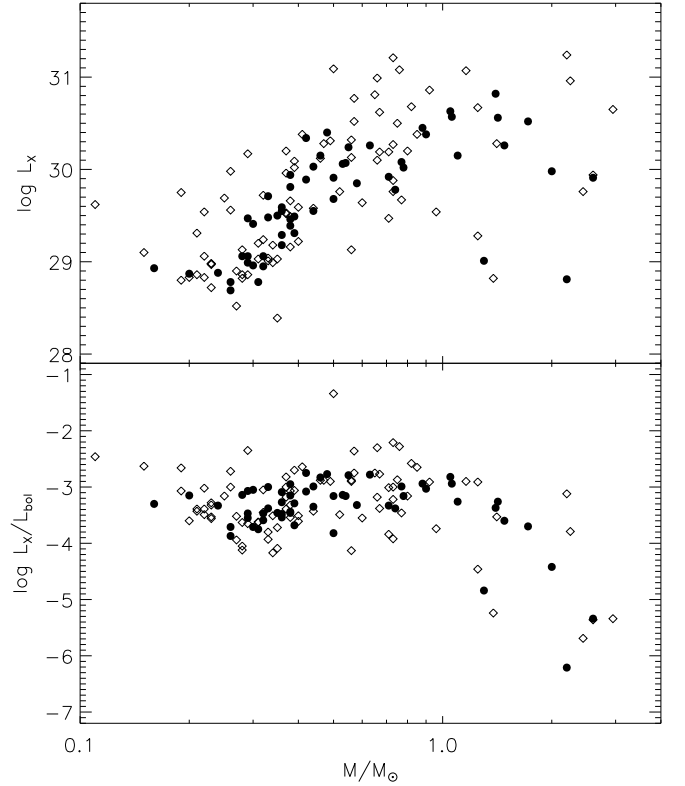}}
\caption{Comparison of $L_\mathrm{X}$ ({\em top}) and
$L_\mathrm{X}/L_\mathrm{bol}$ ({\em bottom} vs. mass for detected members
and candidates of the $\lambda$~Ori (filled circles) and $\sigma$~Ori (open
diamonds) clusters.} 
\label{lx-mass_lsori}
\end{figure}

\section{Comparison with the $\sigma$~Ori cluster}
\label{sigori}

In this section we compare our results with those obtained for the
$\sigma$~Ori cluster. This cluster is very similar to $\lambda$~Ori, with a
few hundred PMS stars concentrated around a central O star and a comparable
age of $\sim$\,2$-$5~Myr. However, as mentioned in Sect.~\ref{intro}, the
two clusters differ for the higher fraction of stars with discs and stars
with active accretion observed in $\sigma$~Ori \citep[and references
therein]{sacco08} and for the supernova explosion that affected the
$\lambda$~Ori region. 

The $\sigma$~Ori cluster was observed by {\em XMM-Newton} by FPS06 at a
similar sensitivity, and the data were analysed consistently with our
present analysis, so that the results of the two observations can be
directly compared. Another {\em XMM-Newton} observation to the west of the
cluster centre has been performed by \citet{lopez08}; however, to avoid
possible biases, we did not include their data in our comparison, since
their observation was performed with a different filter, implying a
different sensitivity limit, and it was analysed using a different method.
After the FPS06 paper was published, several new studies of the $\sigma$~Ori
cluster have become available, which have significantly improved the
membership information for many stars. Before performing the comparison, we
have therefore updated the catalogue of FPS06, using mainly information from
the Mayrit catalogue by \citet{caballero08mayrit}, the spectroscopic studies
by \citet{sacco08} and \citet{maxted08}, the {\em Spitzer} study by
\citet{hernandez07}, the proper motion study by \citet{caballero10pm}, and
updated photometry from \citet{kenyon05} and \citet{mayne08}. We rejected
$\sim$\,20\% (40/210) of the late-type stars considered as members by FPS06,
most of which were not detected in X-rays, and added 25 new members and
candidates with $M\sim\,0.1-2\,M_\odot$, nine of which were detected by
FPS06; for the others, upper limits were computed as described in FPS06 and
in Sect.~\ref{ident}. Masses for $\sigma$~Ori members were computed in the
same way as for $\lambda$~Ori (see Sect.~\ref{opt_cat}). For consistency
with FPS06 and other studies of the $\sigma$~Ori cluster, we adopt the {\em
Hipparcos} distance of 352~pc, although recent determinations give values
between $\sim$\,385~pc and 420~pc \citep{caballero08sigab,mayne08,sherry08}.
Using an average value of 400~pc would only increase the X-ray luminosities
of $\sigma$~Ori stars by $\sim$\,0.1~dex, without significantly affecting
the distribution of stars in the mass bins considered in our analysis.

Figure~\ref{lx-mass_lsori} shows $L_\mathrm{X}$ and
$L_\mathrm{X}/L_\mathrm{bol}$ as a function of mass for detected members of
the $\lambda$~Ori and $\sigma$~Ori clusters. There are no significant
differences in the trends observed for the two clusters, except for a wider
spread in the X-ray luminosities observed in $\sigma$~Ori, which is
particularly evident for masses between $\sim$\,0.5 and $\sim\,1\,M_\odot$:
while the luminosities of $\lambda$~Ori stars differ by less than 1~dex, the
spread for the $\sigma$~Ori cluster is $\sim$\,2~dex. In particular, there
are several $\sigma$~Ori members with $\log L_\mathrm{X} >
30.6$~erg~s$^{-1}$, i.e. brighter than $\lambda$~Ori stars of similar mass.
Most of these objects are very active, displaying strong flares or
significant variability, and hotter plasma than observed in $\lambda$~Ori,
with average temperature above 2~keV. Such high-activity stars are not
observed in $\lambda$~Ori. However, since the number of stars in the
$\lambda$~Ori sample is much lower than in the $\sigma$~Ori one, it is
possible that the lack of stars at high-activity levels might be due to a
statistical bias due to the smallness of the sample, rather than indicating
a difference in magnetic activity between the two clusters.

\begin{figure}
\resizebox{\hsize}{!}{\includegraphics[clip]{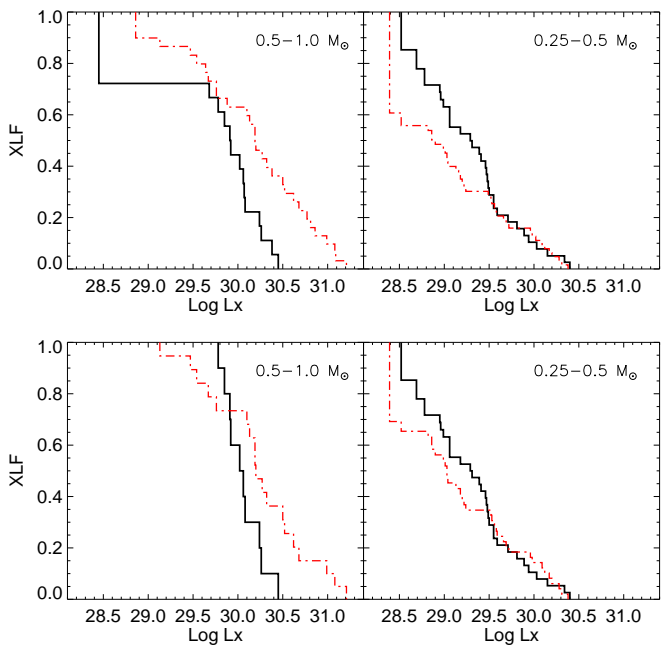}}
\caption{{\em Upper panels:} comparison between the XLFs of the
$\lambda$~Ori (black solid line) and $\sigma$~Ori (red dash-dotted line)
clusters for members and candidates with masses 0.5$-$1\,$M_\odot$ ({\em
left}) and 0.25$-$0.5\,$M_\odot$ ({\em right}). {\em Lower panels}: same
comparison only for spectroscopically-confirmed members.}
\label{xldf_lsori}
\end{figure}

\begin{table}
\centering
\caption{\label{medians}
Comparison between the luminosity distributions of the $\lambda$~Ori and
$\sigma$~Ori clusters.}
\begin{tabular}{lccc}
\hline\hline
Mass bin & \multicolumn{2}{c}{Medians}& Probability\tablefootmark{a}\\
 & $\lambda$~Ori& $\sigma$~Ori& \\
\hline
\multicolumn{4}{c}{All members and candidates}\\
$0.5-1.0\,M_\odot$ & 29.9& 30.2& 0.04$-$0.09\\
$0.25-0.5\,M_\odot$& 29.3& 28.9& 0.03$-$0.09\\
\hline
\multicolumn{4}{c}{Spectroscopically-confirmed members}\\
$0.5-1.0\,M_\odot$ & 30.0& 30.2& 0.03$-$0.95\\
$0.25-0.5\,M_\odot$& 29.3& 29.0& 0.20$-$0.45\\
\hline
\end{tabular}
\tablefoot{
\tablefoottext{a} Range of probabilities that the two samples are drawn from
the same parent population, obtained from the two-sample tests in
\sc{asurv}.
} 
\end{table}

In Fig.~\ref{xldf_lsori} we compare the XLFs of $\lambda$~Ori and
$\sigma$~Ori stars in the 0.5$-$1\,$M_\odot$ and 0.25$-$0.5\,$M_\odot$ mass
bins, and the medians of the distributions are given in Table~\ref{medians}.
In the upper panels, we show the comparison for the whole sample of cluster
members and candidates. As already suggested by Fig.~\ref{lx-mass_lsori},
stars with $M=0.5-1\,M_\odot$ in $\lambda$~Ori are less luminous in X-rays
than those of $\sigma$~Ori, with median X-ray luminosities differing by a
factor of 2. The two-sample tests in {\sc asurv} confirm that the two
distributions differ at the 90\% level. On the other hand, stars with
$M=0.25-0.5\,M_\odot$ appear to be more luminous in $\lambda$~Ori, although
the high-luminosity tails of the two XLFs are comparable. 

The difference between the XLFs of stars with $M=0.5-1\,M_\odot$ can be 
explained, at least in part, with contamination of the samples by
non-members. As mentioned in Sect.~\ref{xlum}, all the non-detections among
$\lambda$~Ori stars in this mass range are only photometric candidates,
which are likely to be older field stars, rather than true cluster members.
Including these contaminants as candidates results in a lower median
luminosity of the $\lambda$~Ori sample. On the other hand, as shown in
Fig.~\ref{lx-mass_lsori}, in the $\sigma$~Ori sample there are some
high-luminosity active stars that are not observed in $\lambda$~Ori. Half of
these bright stars  are young objects with unknown radial velocity, and it
is possible that some of them might belong to one of the foreground young
populations that are present in the region
\citep[e.g.][]{alcala00,jeffries06}. To exclude the effects of contamination
by non-members, in the bottom panel of Fig.~\ref{xldf_lsori} we compare the
XLFs computed considering only spectroscopically-confirmed members, i.e.
stars having both signatures of youth and radial velocity consistent with
that of the clusters. In the 0.5$-$1\,$M_\odot$ mass bin, all confirmed
members in $\lambda$~Ori and 95\% of those in $\sigma$~Ori are detected, and
the discrepancy between the two XLFs is reduced, although the medians still
differ by 0.2~dex. Similar results are obtained for the lower mass bin. In
both cases, the results of the two-sample tests do not allow us to reject
the hypothesis that the two distributions are derived from the same parent
population. As a result, the differences observed in the upper panels of
Fig.~\ref{xldf_lsori} are not real, but can be attributed to the
uncertainties in the membership information of the two clusters. The
residual differences in the XLFs of confirmed members are comparable to the
expected uncertainties in the derived X-ray luminosities due to the
uncertainties in the distance of $\sigma$~Ori and in the derived conversion
factors. Therefore, we conclude that there is no convincing evidence for any
significant difference in the X-ray properties of the two clusters.

\section{Summary and conclusions}
\label{concl}

In this paper we have presented the analysis of an {\em XMM-Newton}
observation of the $\lambda$~Ori cluster, centred on the O8~III star
$\lambda$~Ori\,AB. We derived the X-ray properties of cluster members and
analysed the EPIC spectra of the central hot star and of the brightest
sources in the field. Our results can be summarised as follows.

\begin{itemize}
\item We detected 58 cluster members and candidates above a limiting
sensitivity of $\sim\,3\times 10^{28}$~erg~s$^{-1}$, spanning the cluster
sequence from the central O8III star $\lambda$~Ori\,AB down to stars with
$M\sim\,0.2\,M_\odot$, and identified 24 new possible cluster candidates
from the 2MASS catalogue. We did not detect any X-ray emission from 12
Class~III candidate members with masses between 0.4 and $2\,M_\odot$,
suggesting that these objects might be older field stars rather than cluster
members.

\item We found significant variability in $\sim$\,35\% of detected members
and candidates, including strong flares in $\sim$\,10\% of them, in
agreement with other observations of SFRs and very young clusters.

\item The emission from the central O\,III star is soft, with the bulk of
the plasma at temperatures of 0.2$-$0.3~keV, and an X-ray luminosity of
$\sim 10^{32}$~erg~s$^{-1}$, in agreement with other observations of massive
stars, and consistently with a wind origin. We did not find the hot
component at $\sim$\,25~MK derived from {\em ASCA}, which can be attributed
to contamination by the sources surrounding $\lambda$~Ori\,AB and included
in the large {\em ASCA} extraction region. On the other hand, late-type
stars show harder spectra, with coronal temperatures of 0.3$-$0.9~keV and
0.8$-$2.7~keV and strongly subsolar abundances, as commonly observed in PMS
late-type stars.

\item The high X-ray luminosity derived from spectral analysis of the
weak-lined T~Tauri star HD\,245059 confirms that it does not belong to the
$\lambda$~Ori cluster, but it is likely a foreground PMS star.

\item We found that $L_\mathrm{X}$ increases with stellar mass up to
2\,$M_\odot$, with a slope of $\sim$\,1.9, in agreement with the results
obtained for other SFRs and young clusters, while $\log
L_\mathrm{X}/L_\mathrm{bol}$ is nearly constant around a median value of
$-3.5$. No significant difference is found in the X-ray luminosity of stars
with or without circustellar discs.

\item Finally, we compared the X-ray properties of $\lambda$~Ori late-type
stars with those of the coeval $\sigma$~Ori cluster. The properties of the
two clusters appear to be very similar, suggesting that stellar activity in
the $\lambda$~Ori cluster has not been significantly affected by the
different ambient environment.

\end{itemize}

\begin{acknowledgements}
We thank J.~A.~Caballero for his useful comments as referee of this paper.
We acknowledge partial financial support from Ministero dell'Istruzione,
Universit\`a e Ricerca (MIUR). Research on X-rays from young stars by G.S.
is supported by NASA/Goddard XMM-Newton Guest Observer Facility grants
NNX09AT15G and NNX09AC11G and NASA Astrophysics Data Analysis programme
grant NNX09AC96G to RIT.
This publication makes use of data products from the Two Micron All Sky
Survey, which is a joint project of the University of Massachusetts and the
Infrared Processing and Analysis Center/California Institute of Technology,
and of the Guide Star Catalogue-II, which is a joint project of the Space
Telescope Science Institute and the Osservatorio Astronomico di Torino. This
research has made use of the Simbad and VizieR databases available at the
CDS.
\end{acknowledgements}

\bibliographystyle{aa} 
\bibliography{biblio}

\appendix
\section{X-ray detections and upper limits}

\longtab{1}{
\tiny 
\begin{longtable}{rccrrrlcl}
\caption{\label{det_all}
X-ray sources detected in the $\lambda$~Ori cluster.}\\
\hline\hline
LOX& RA$_X$& DEC$_X$& Offaxis& Sign.$^a$& Count rate$^b$& Identification$^c$& Offset& Notes$^d$ \\
   &\multicolumn{2}{c}{(J2000)}& (arcmin)& & (cts/ks)& & (arcsec)& \\
\hline
\endfirsthead
\caption{Continued.}\\
\hline\hline
LOX& RA$_X$& DEC$_X$& Offaxis& Sign.$^a$& Count rate$^b$& Identification$^c$& Offset& Notes$^d$ \\
   &\multicolumn{2}{c}{(J2000)}& (arcmin)& & (cts/ks)& & (arcsec)& \\
\hline
\endhead
\hline
\endfoot
\hline
\multicolumn{9}{l}{\begin{minipage}{0.86\textwidth}
{\bf Notes.} 
$^{(a)}$ Detection significance. 
$^{(b)}$ MOS-equivalent count rates in the 0.3$-$7.8~keV band. 
$^{(c)}$ Identifications labelled DM, LOri and Lori-SOC are from
\citet{dm99}, \citet{barrado04}, and \citet{bouy09lam}, respectively.
2MASS~J05340691+1001005 (LOX\,1) and J05351974+0947476 (LOX\,99) are stars
X2 and X4 from \citet{stone85}. The two stars with TYC identification are
stars b and f in \citet{murdin77}.
$^{(d)}$ Member = previously known members and candidates, NM = previously
known non-members, New?= possible new candidates. For known non-members we
also indicate in parenthesis the reason for membership exclusion. 
$^{(e)}$ This star has a fainter visual companion (2MASS~J05340664+1001034,
$J = 13.10$~mag) at a distance of $\sim$\,5~arcsec, which however falls
outside our identification radius (offset = 4.2~arcsec).
\end{minipage}}\\
\endlastfoot
  1& 05:34:06.86& $+$10:01:00.8& 15.94&  39.4&  26.69 $\pm$ 1.31& 2MASS J05340691+1001005$^e$& 0.94& Member \\
  2& 05:34:08.35& $+$09:51:25.5& 15.50&   8.4&   2.28 $\pm$ 0.41& LOri 044               & 0.62  & NM (no $v_\mathrm{r}$)\\
  3& 05:34:11.83& $+$09:57:03.2& 13.97&  10.0&   2.24 $\pm$ 0.33& LOri 052               & 0.74  & NM (no $v_\mathrm{r}$)\\
  4& 05:34:18.31& $+$09:52:37.4& 12.80&  19.2&   3.81 $\pm$ 0.35& 2MASS J05342809+0948476& 0.54  & New?   \\
  5& 05:34:18.42& $+$09:58:04.8& 12.47&   6.8&   0.94 $\pm$ 0.19& \ldots                 & \ldots&        \\
  6& 05:34:24.63& $+$09:57:01.2& 10.82&   9.8&   1.40 $\pm$ 0.21& \ldots                 & \ldots&        \\
  7& 05:34:26.09& $+$09:51:47.4& 11.26&   6.5&   0.86 $\pm$ 0.17& LOri 046               & 2.02  & NM (no $v_\mathrm{r}$, Li)\\
  8& 05:34:28.09& $+$09:48:45.9& 12.32&  18.6&   3.57 $\pm$ 0.33& 2MASS J05342809+0948476& 1.71  & New?   \\
  9& 05:34:29.76& $+$09:51:33.5& 10.52&   6.7&   1.09 $\pm$ 0.21& 2MASS J05342960+0951317& 2.91  &        \\
 10& 05:34:31.98& $+$09:56:29.8&  8.98&   5.0&   0.39 $\pm$ 0.10& GSC2.3 N9O7015492      & 1.89  &        \\
 11& 05:34:32.72& $+$09:43:06.7& 15.64&  14.9&   5.63 $\pm$ 0.64& \ldots                 & \ldots&        \\
 12& 05:34:32.77& $+$09:59:31.7&  9.44&  43.3&   8.12 $\pm$ 0.39& DM 9                   & 1.18  & Member \\
 13& 05:34:34.43& $+$10:03:17.8& 11.06&   8.2&   1.94 $\pm$ 0.30& \ldots                 & \ldots&        \\
 14& 05:34:34.80& $+$10:07:04.6& 13.78& 383.4& 628.79 $\pm$ 4.71& HD 245059              & 2.35  & NM (no ph., $v_\mathrm{r}$, $\mu$) \\
 15& 05:34:35.38& $+$09:47:07.3& 12.08&   6.5&   1.56 $\pm$ 0.31& \ldots                 & \ldots&        \\
 16& 05:34:35.56& $+$09:59:44.0&  8.88&  30.0&   4.66 $\pm$ 0.29& DM 11                  & 1.07  & Member \\
 17& 05:34:36.20& $+$09:53:44.7&  8.26&  62.1&  13.85 $\pm$ 0.49& DM 12                  & 0.63  & Member \\
 18& 05:34:36.30& $+$10:03:46.0& 11.04&   5.2&   0.50 $\pm$ 0.13& USNO-B1.0 1000-0060503 & 2.43  &        \\
 19& 05:34:39.18& $+$09:52:55.6&  7.85&  34.1&   5.62 $\pm$ 0.31& DM 14                  & 0.50  & Member \\
 20& 05:34:39.27& $+$10:01:30.4&  9.01&  24.3&   3.38 $\pm$ 0.25& LOri 036               & 1.68  & NM (no $v_\mathrm{r}$, Li) \\
 21& 05:34:39.74& $+$10:06:21.7& 12.49&  29.1&   6.46 $\pm$ 0.41& DM 16                  & 0.93  & Member \\
 22& 05:34:42.86& $+$09:51:58.6&  7.50&   5.5&   0.33 $\pm$ 0.09& \ldots                 & \ldots&        \\
 23& 05:34:44.75& $+$10:05:47.0& 11.34&   6.4&   1.09 $\pm$ 0.20& USNO-B1.0 1000-0060556 & 3.53  &        \\
 24& 05:34:45.43& $+$09:59:21.7&  6.55&  23.4&   3.45 $\pm$ 0.28& \ldots                 & \ldots&        \\
 25& 05:34:45.46& $+$10:01:08.1&  7.60&   9.8&   0.94 $\pm$ 0.15& \ldots                 & \ldots&        \\
 26& 05:34:47.24& $+$10:02:43.2&  8.46&  75.5&  21.20 $\pm$ 0.63& DM 18                  & 0.19  & Member \\
 27& 05:34:48.17& $+$09:43:26.0& 13.57&   7.6&   1.75 $\pm$ 0.30& LOri 068               & 2.16  & Member \\
 28& 05:34:48.26& $+$09:52:44.0&  5.97&  10.0&   0.65 $\pm$ 0.10& \ldots                 & \ldots&        \\
 29& 05:34:48.43& $+$09:57:15.4&  5.06&  46.0&   5.63 $\pm$ 0.26& DM 19                  & 0.53  & Member \\
 30& 05:34:49.00& $+$09:58:02.8&  5.18&  21.7&   2.00 $\pm$ 0.16& DM 20                  & 0.82  & NM (no $v_\mathrm{r}$) \\
 31& 05:34:50.46& $+$09:51:47.7&  6.14&  66.1&  11.36 $\pm$ 0.39& DM 22                  & 0.17  & Member \\
 32& 05:34:50.84& $+$10:04:30.5&  9.50&  27.3&   4.59 $\pm$ 0.32& 4C 09.21               & 0.35  & Radio source\\
 33& 05:34:51.72& $+$09:45:58.9& 10.88&  12.7&   2.21 $\pm$ 0.26& \ldots                 & \ldots&        \\
 34& 05:34:52.63& $+$09:55:50.4&  3.89&  16.7&   1.02 $\pm$ 0.11& 2MASS J05345260+0955500& 0.48  & New?   \\
 35& 05:34:53.21& $+$09:42:41.6& 13.87&   8.5&   0.66 $\pm$ 0.15& \ldots                 & \ldots&        \\
 36& 05:34:54.79& $+$09:53:34.3&  4.17&  39.7&   4.45 $\pm$ 0.23& \ldots                 & \ldots&        \\
 37& 05:34:55.25& $+$10:00:34.3&  5.56&   7.5&   0.29 $\pm$ 0.06& LOri 075               & 0.53  & Member \\
 38& 05:34:55.59& $+$09:56:09.7&  3.16&   6.6&   0.44 $\pm$ 0.09& LOri-SOC-1             & 0.62  & Member \\
 39& 05:34:55.66& $+$09:57:57.4&  3.67&   8.3&   0.51 $\pm$ 0.09& 2MASS J05345564+0957581& 0.82  & New?   \\
 40& 05:34:55.92& $+$09:58:43.2&  4.07&  14.8&   1.10 $\pm$ 0.12& \ldots                 & \ldots&        \\
 41& 05:34:56.44& $+$09:55:04.8&  3.10&  14.3&   1.16 $\pm$ 0.14& LOri 050               & 0.57  & Member \\
 42& 05:34:56.71& $+$09:54:54.3&  3.10&   5.1&   0.51 $\pm$ 0.18& LOri-SOC-2             & 0.70  & Member \\
 43& 05:34:56.98& $+$09:57:27.0&  3.14&   9.0&   0.57 $\pm$ 0.09& \ldots                 & \ldots&        \\
 44& 05:34:57.11& $+$09:54:37.0&  3.13&  40.8&   4.73 $\pm$ 0.23& LOri 024               & 0.34  & Member \\
 45& 05:34:57.59& $+$09:46:07.1& 10.29&  34.9&   8.19 $\pm$ 0.43& LOri 020               & 0.28  & NM (no $v_\mathrm{r}$, Li) \\
 46& 05:34:58.20& $+$09:56:27.4&  2.54&  46.3&   5.24 $\pm$ 0.23& HD 245140              & 0.78  & Member \\
 47& 05:34:58.32& $+$09:53:47.0&  3.36&   7.1&   0.55 $\pm$ 0.10& LOri 056               & 0.77  & Member \\
 48& 05:34:58.75& $+$09:47:29.1&  8.89&  10.2&   1.55 $\pm$ 0.21& \ldots                 & \ldots&        \\
 49& 05:34:59.06& $+$10:02:07.5&  6.49&   8.5&   0.71 $\pm$ 0.12& \ldots                 & \ldots&        \\
 50& 05:34:59.11& $+$10:06:01.0& 10.22&   5.6&   0.80 $\pm$ 0.18& \ldots                 & \ldots&        \\
 51& 05:34:59.40& $+$09:53:12.3&  3.61&  30.3&   3.21 $\pm$ 0.20& \ldots                 & \ldots&        \\
 52& 05:35:00.79& $+$09:51:52.2&  4.58&  12.8&   1.15 $\pm$ 0.15& 2MASS J05350064+0951510& 2.40  & New?   \\
 53& 05:35:01.47& $+$09:47:53.5&  8.34&  11.1&   1.20 $\pm$ 0.16& \ldots                 & \ldots&        \\
 54& 05:35:01.85& $+$09:47:21.5&  8.84&   5.0&   0.88 $\pm$ 0.31& \ldots                 & \ldots&        \\
 55& 05:35:02.66& $+$09:56:48.0&  1.60&  24.1&   1.74 $\pm$ 0.13& LOri 043               & 1.21  & Member \\
 56& 05:35:03.07& $+$09:56:16.8&  1.33&  27.5&   2.68 $\pm$ 0.17& 2MASS J05350309+0956162& 0.67  & New?   \\
 57& 05:35:03.24& $+$10:02:52.2&  6.93&   7.9&   0.68 $\pm$ 0.13& 2MASS J05350327+1002532& 1.19  &        \\
 58& 05:35:03.59& $+$09:50:53.7&  5.29&  16.2&   1.37 $\pm$ 0.14& 2MASS J05350356+0950531& 0.68  & New?   \\
 59& 05:35:05.03& $+$09:56:55.9&  1.21&  14.0&   1.29 $\pm$ 0.15& 2MASS J05350496+0956561& 0.88  & New?   \\
 60& 05:35:05.22& $+$09:55:15.8&  1.11&  44.5&   7.77 $\pm$ 0.39& 2MASS J05350528+0955149& 1.31  & New?   \\
 61& 05:35:05.48& $+$09:42:46.7& 13.29&   9.0&   1.57 $\pm$ 0.27& \ldots                 & \ldots&        \\
 62& 05:35:05.99& $+$10:00:19.2&  4.31&  11.1&   0.61 $\pm$ 0.09& DM 24                  & 1.11  & Member \\
 63& 05:35:06.04& $+$10:00:36.3&  4.59&   6.6&   0.32 $\pm$ 0.08& GSC2.3 N9O4013323      & 0.25  &        \\
 64& 05:35:06.47& $+$09:59:58.5&  3.95&  11.9&   0.68 $\pm$ 0.09& TYC 705-860-1          & 1.38  & NM (no ph.) \\
 65& 05:35:06.80& $+$09:57:30.2&  1.50&   6.1&   0.36 $\pm$ 0.09& LOri 066               & 2.44  & Member \\
 66& 05:35:06.97& $+$09:48:57.2&  7.11&  29.8&   3.63 $\pm$ 0.24& DM 25                  & 0.52  & Member \\
 67& 05:35:07.45& $+$09:58:22.7&  2.34&  46.9&   6.56 $\pm$ 0.26& LOri 045               & 0.52  & Member \\
 68& 05:35:07.88& $+$09:51:46.0&  4.29&   8.2&   0.32 $\pm$ 0.07& \ldots                 & \ldots&        \\
 69& 05:35:07.99& $+$09:50:55.9&  5.12&  16.2&   1.11 $\pm$ 0.12& 2MASS J05350794+0950545& 1.50  & New?   \\
 70& 05:35:08.15& $+$09:55:34.1&  0.49&  45.2&   5.77 $\pm$ 0.27& $\lambda$~Ori\,C       & 0.29  & Member \\
 " &            &              &      &      &                  & LOri-MAD-30            & 0.25  &        \\
 71& 05:35:08.25& $+$09:56:03.0&  0.04& 691.6& 574.70 $\pm$ 2.21& $\lambda$ Ori\,AB      & 1.01  & Member \\
 72& 05:35:08.31& $+$09:57:23.5&  1.34&  41.1&   4.45 $\pm$ 0.21& \ldots                 & \ldots&        \\
 73& 05:35:08.33& $+$09:42:53.6& 13.16&  64.9&  27.59 $\pm$ 0.91& DM 26                  & 0.25  & Member \\
 74& 05:35:09.28& $+$10:04:56.1&  8.88&   5.0&   0.28 $\pm$ 0.10& \ldots                 & \ldots&        \\
 75& 05:35:09.60& $+$10:01:50.5&  5.80&   8.2&   0.39 $\pm$ 0.08& HD245185               & 1.03  & Member \\
 76& 05:35:10.00& $+$09:50:33.8&  5.50&  23.1&   2.50 $\pm$ 0.22& 2MASS J05351006+0950328& 1.44  & New?   \\
 77& 05:35:10.22& $+$09:53:36.9&  2.48&   6.5&   0.34 $\pm$ 0.08& \ldots                 & \ldots&        \\
 78& 05:35:11.35& $+$10:00:51.4&  4.86&   7.4&   0.69 $\pm$ 0.12& LOri 057               & 1.29  & Member \\
 79& 05:35:12.14& $+$09:55:21.7&  1.15&  50.6&   7.56 $\pm$ 0.30& 2MASS J05351205+0955218& 1.14  & New?   \\
 80& 05:35:12.46& $+$09:53:09.2&  3.07&  24.0&   2.13 $\pm$ 0.15& LOri 048               & 2.40  & Member \\
 81& 05:35:13.47& $+$09:55:25.3&  1.40&  44.5&   4.97 $\pm$ 0.23& LOri 016               & 0.09  & Member \\
 82& 05:35:13.65& $+$09:56:27.2&  1.35&  50.2&   7.10 $\pm$ 0.26& LOri 019               & 0.08  & Member \\
 83& 05:35:14.60& $+$09:50:03.7&  6.18&  14.3&   1.31 $\pm$ 0.14& 2MASS J05351456+0950026& 1.19  & New?   \\
 84& 05:35:15.12& $+$10:01:06.7&  5.32&  19.2&   1.23 $\pm$ 0.12& DM 29                  & 0.28  & Member \\
 85& 05:35:15.44& $+$09:48:37.0&  7.64&   5.5&   0.36 $\pm$ 0.09& LOri 062               & 1.67  & Member \\
 86& 05:35:16.02& $+$09:53:37.8&  3.06&  29.4&   2.91 $\pm$ 0.18& 2MASS J05351606+0953374& 0.71  & New?   \\
 87& 05:35:16.21& $+$09:55:20.0&  2.05&  75.4&  13.25 $\pm$ 0.38& LOri 006               & 1.55  & Member \\
 88& 05:35:16.99& $+$10:10:16.9& 14.39&   6.9&   1.77 $\pm$ 0.45& \ldots                 & \ldots&        \\
 89& 05:35:17.16& $+$09:51:12.0&  5.31&  23.5&   2.12 $\pm$ 0.17& DM 30                  & 0.57  & Member \\
 90& 05:35:17.29& $+$09:49:24.9&  6.99&   8.6&   1.07 $\pm$ 0.17& \ldots                 & \ldots&        \\
 91& 05:35:17.88& $+$09:56:59.0&  2.51&   8.1&   0.59 $\pm$ 0.09& LOri 065               & 1.87  & Member \\
 92& 05:35:17.89& $+$09:54:16.6&  2.94&  30.9&   3.67 $\pm$ 0.23& 2MASS J05351794+0954167& 0.67  & New?   \\
 93& 05:35:18.13& $+$09:52:24.1&  4.37&   9.6&   0.69 $\pm$ 0.13& LOri 061               & 0.79  & Member \\
 94& 05:35:18.26& $+$10:02:38.6&  7.02&  18.2&   1.84 $\pm$ 0.22& DM 32                  & 1.72  & Member \\
 95& 05:35:18.48& $+$09:57:37.4&  2.94&   6.8&   0.53 $\pm$ 0.12& \ldots                 & \ldots&        \\
 96& 05:35:18.81& $+$09:44:04.2& 12.25&   6.0&   2.13 $\pm$ 0.46& 2MASS J05351857+0944058& 2.53  & New?   \\
 97& 05:35:19.13& $+$09:53:57.4&  3.37&  12.2&   0.76 $\pm$ 0.11& \ldots                 & \ldots&        \\
 98& 05:35:19.14& $+$09:54:55.1&  2.88& 190.5&  61.84 $\pm$ 0.92& 2MASS J05351857+0944058& 1.33  & New?   \\
 99& 05:35:19.75& $+$09:47:47.2&  8.73& 138.7&  54.48 $\pm$ 0.95& 2MASS J05351974+0947476& 0.37  & Member \\
100& 05:35:19.96& $+$10:02:37.2&  7.16&  43.6&   7.63 $\pm$ 0.35& DM 33                  & 0.91  & Member \\
101& 05:35:20.05& $+$09:50:33.7&  6.19&  16.6&   1.50 $\pm$ 0.18& \ldots                 & \ldots&        \\
102& 05:35:20.07& $+$09:49:05.4&  7.53&   5.9&   0.53 $\pm$ 0.12& LOri 060               & 1.34  & Member \\
103& 05:35:20.54& $+$09:52:57.0&  4.31&   6.0&   0.25 $\pm$ 0.06& 2MASS J05352036+0952546& 3.52  &        \\
104& 05:35:20.90& $+$09:48:25.2&  8.23&   5.1&   0.30 $\pm$ 0.09& \ldots                 & \ldots&        \\
105& 05:35:21.14& $+$10:01:09.7&  6.00&   7.0&   0.47 $\pm$ 0.09& \ldots                 & \ldots&        \\
106& 05:35:21.40& $+$09:49:56.4&  6.90&  21.5&   2.34 $\pm$ 0.19& LOri 055               & 0.49  & Member \\
107& 05:35:21.42& $+$09:54:55.0&  3.40&   7.6&   0.39 $\pm$ 0.09& 2MASS J05352135+0954549& 1.03  & New?   \\
108& 05:35:21.44& $+$09:44:09.3& 12.32&  11.6&   1.46 $\pm$ 0.21& DM 34                  & 1.18  & Member \\
109& 05:35:22.16& $+$09:53:59.2&  3.97&  38.3&   4.20 $\pm$ 0.21& 2MASS J05352216+0953586& 0.50  & New?   \\
110& 05:35:22.19& $+$09:52:27.1&  4.95&  53.6&   7.08 $\pm$ 0.28& DM 35                  & 0.41  & Member \\
111& 05:35:22.39& $+$09:50:05.6&  6.88&   5.1&   0.62 $\pm$ 0.20& \ldots                 & \ldots&        \\
112& 05:35:23.24& $+$09:52:18.2&  5.24&   6.2&   0.18 $\pm$ 0.06& 2MASS J05352320+0952190& 0.95  & New?   \\
113& 05:35:23.56& $+$09:57:57.2&  4.19&  17.2&   1.16 $\pm$ 0.12& TYC 705-937-1          & 0.60  & NM (no ph.) \\
114& 05:35:25.32& $+$10:08:38.7& 13.26&  21.7&  13.03 $\pm$ 1.09& DM 36                  & 0.86  & Member \\
115& 05:35:25.43& $+$09:47:40.3&  9.37&  13.9&   1.67 $\pm$ 0.19& USNO-B1.0 0997-0080734 & 0.45  &        \\
116& 05:35:25.48& $+$09:53:03.0&  5.17&  10.1&   0.60 $\pm$ 0.10& \ldots                 & \ldots&        \\
117& 05:35:27.17& $+$09:53:11.1&  5.44&  47.4&   6.08 $\pm$ 0.27& GSC2.3 N9O7021134      & 0.18  &        \\
118& 05:35:27.25& $+$10:00:12.5&  6.23&  10.2&   0.72 $\pm$ 0.11& \ldots                 & \ldots&        \\
119& 05:35:27.49& $+$09:43:53.9& 13.03&   6.5&   0.97 $\pm$ 0.20& \ldots                 & \ldots&        \\
120& 05:35:27.77& $+$09:56:04.1&  4.77&   8.0&   0.51 $\pm$ 0.09& GSC2.3 N9O7021459      & 0.71  &        \\
121& 05:35:28.41& $+$10:02:28.3&  8.09&   8.2&   0.58 $\pm$ 0.11& 2MASS J05352846+1002275& 1.09  & New?   \\
122& 05:35:28.49& $+$10:03:11.1&  8.68&   6.4&   0.48 $\pm$ 0.10& \ldots                 & \ldots&        \\
123& 05:35:29.32& $+$09:46:32.2& 10.82&  20.5&   3.19 $\pm$ 0.27& 2MASS J05352920+0946317& 1.67  & New?   \\
124& 05:35:29.73& $+$09:56:27.9&  5.27&  18.7&   1.36 $\pm$ 0.14& \ldots                 & \ldots&        \\
125& 05:35:30.02& $+$09:59:26.8&  6.31&   9.3&   0.68 $\pm$ 0.10& LOri 080               & 1.39  & Member \\
126& 05:35:30.33& $+$09:47:54.7&  9.77&   7.9&   1.08 $\pm$ 0.20& \ldots                 & \ldots&        \\
127& 05:35:30.36& $+$10:00:01.6&  6.71&   6.7&   0.35 $\pm$ 0.07& \ldots                 & \ldots&        \\
128& 05:35:30.46& $+$09:50:33.3&  7.73&  17.2&   1.90 $\pm$ 0.23& DM 38                  & 0.83  & Member \\
129& 05:35:30.60& $+$09:54:30.2&  5.68&  29.6&   3.26 $\pm$ 0.22& \ldots                 & \ldots&        \\
130& 05:35:30.76& $+$09:56:10.9&  5.51&  10.6&   0.63 $\pm$ 0.10& USNO-B1.0 0999-0067187 & 3.77  &        \\
131& 05:35:32.49& $+$09:57:54.7&  6.22&  24.7&   2.87 $\pm$ 0.20& \ldots                 & \ldots&        \\
132& 05:35:32.55& $+$09:56:13.9&  5.95&  11.5&   0.87 $\pm$ 0.12& \ldots                 & \ldots&        \\
133& 05:35:33.40& $+$09:51:47.7&  7.49&  22.6&   2.39 $\pm$ 0.19& \ldots                 & \ldots&        \\
134& 05:35:33.66& $+$09:46:28.4& 11.42&   5.7&   0.97 $\pm$ 0.20& \ldots                 & \ldots&        \\
135& 05:35:34.78& $+$10:00:35.8&  7.93&  60.6&  10.46 $\pm$ 0.37& DM 39                  & 0.99  & Member \\
136& 05:35:35.41& $+$09:54:43.1&  6.79&   7.1&   0.41 $\pm$ 0.08& USNO-B1.0 0999-0067223 & 2.52  &        \\
137& 05:35:35.87& $+$09:44:36.3& 13.30&   7.1&   0.91 $\pm$ 0.22& DM 40                  & 2.44  & Member \\
138& 05:35:35.96& $+$09:47:51.8& 10.64&   6.3&   0.54 $\pm$ 0.14& GSC2.3 N9O7020712      & 0.90  &        \\
139& 05:35:36.02& $+$09:56:48.8&  6.84&   5.4&   0.58 $\pm$ 0.14& \ldots                 & \ldots&        \\
140& 05:35:37.98& $+$09:44:07.6& 13.98&  22.3&   6.32 $\pm$ 0.50& GSC2.3 N9O7020552      & 0.15  &        \\
141& 05:35:38.14& $+$09:53:16.2&  7.83&  19.5&   2.00 $\pm$ 0.18& 2MASS J05353811+0953163& 0.31  &        \\
142& 05:35:38.21& $+$10:01:03.8&  8.89&  19.4&   2.13 $\pm$ 0.19& \ldots                 & \ldots&        \\
143& 05:35:39.02& $+$09:55:54.4&  7.54&   8.2&   0.63 $\pm$ 0.11& \ldots                 & \ldots&        \\
144& 05:35:39.53& $+$09:50:33.1&  9.44&  24.9&   3.86 $\pm$ 0.28& DM 41                  & 0.79  & Member \\
145& 05:35:41.35& $+$09:53:55.4&  8.39&   8.5&   0.72 $\pm$ 0.12& \ldots                 & \ldots&        \\
146& 05:35:43.24& $+$09:59:56.2&  9.42&  46.6&   8.23 $\pm$ 0.37& \ldots                 & \ldots&        \\
147& 05:35:43.49& $+$09:54:25.8&  8.79&   5.7&   0.45 $\pm$ 0.11& LOri 083               & 1.60  & Member \\
148& 05:35:43.91& $+$09:56:01.7&  8.74&  10.8&   0.77 $\pm$ 0.12& \ldots                 & \ldots&        \\
149& 05:35:44.61& $+$10:04:50.8& 12.52&   5.5&   0.65 $\pm$ 0.15& \ldots                 & \ldots&        \\
150& 05:35:45.02& $+$09:55:19.7&  9.05&  22.3&   2.78 $\pm$ 0.22& 2MASS J05354495+0955190& 1.12  & New?   \\
 " &            &              &      &      &                  & 2MASS J05354519+0955203& 2.77  & New?   \\
151& 05:35:47.63& $+$09:45:50.8& 14.06&  15.2&   2.84 $\pm$ 0.36& LOri 004               & 1.09  & Member \\
152& 05:35:48.82& $+$09:49:24.4& 11.97&  11.3&   1.46 $\pm$ 0.20& \ldots                 & \ldots&        \\
153& 05:35:49.17& $+$10:00:16.5& 10.89&   8.3&   1.12 $\pm$ 0.18& \ldots                 & \ldots&        \\
154& 05:35:49.60& $+$10:04:33.8& 13.24&  10.7&   2.45 $\pm$ 0.38& \ldots                 & \ldots&        \\
155& 05:35:49.98& $+$09:57:52.1& 10.40&  15.7&   1.82 $\pm$ 0.19& \ldots                 & \ldots&        \\
156& 05:35:51.39& $+$09:55:12.7& 10.62&  30.2&   6.90 $\pm$ 0.44& DM 44                  & 1.67  & Member \\
157& 05:35:51.50& $+$09:54:03.4& 10.80&   5.9&   0.34 $\pm$ 0.09& \ldots                 & \ldots&        \\
158& 05:35:51.73& $+$09:53:34.2& 10.96&  10.6&   1.23 $\pm$ 0.18& 2MASS J05355172+0953340& 0.25  &        \\
159& 05:35:51.93& $+$09:50:28.7& 12.08&  11.4&   1.55 $\pm$ 0.23& LOri 064               & 1.34  & Member \\
160& 05:35:52.51& $+$09:48:33.2& 13.20&  14.6&   3.07 $\pm$ 0.31& LOri 054               & 2.17  & Member \\
161& 05:35:54.32& $+$10:04:23.6& 14.05&  20.0&  10.95 $\pm$ 1.27& DM 45                  & 0.31  & Member \\
162& 05:35:55.46& $+$09:56:30.7& 11.60&  49.6&  23.57 $\pm$ 1.04& DM 46                  & 0.54  & Member \\
163& 05:35:55.74& $+$09:50:52.5& 12.76&   8.8&   1.79 $\pm$ 0.33& DM 47                  & 1.35  & Member \\
164& 05:35:56.78& $+$10:01:52.6& 13.26&   6.6&   1.69 $\pm$ 0.50& \ldots                 & \ldots&        \\
165& 05:35:57.37& $+$09:58:25.9& 12.29&   8.6&   1.43 $\pm$ 0.27& \ldots                 & \ldots&        \\
166& 05:35:58.00& $+$09:54:32.8& 12.31&  41.5&  16.68 $\pm$ 0.90& DM 51                  & 0.36  & Member \\
167& 05:35:59.59& $+$09:50:17.6& 13.86&   9.8&   3.08 $\pm$ 0.67& \ldots                 & \ldots&        \\
\end{longtable} 
}


\longtabL{2}{
\tiny
\begin{landscape}
\begin{longtable}{lccrccrrrrrrrllc}
\caption{\label{upplim}
$3\sigma$ upper limits and optical properties of undetected cluster members
and candidates.}\\
\hline\hline
Identification$^a$ & RA & DEC & Count rate$^b$ & $\log L_\mathrm{X}^c$& $\log L_\mathrm{X}/L_\mathrm{bol}$& 
 $V$& $R$& $I$& $J$& $H$& $K_\mathrm{s}$& Mass & Sp.T& Disc& Mem.$^e$\\
   &\multicolumn{2}{c}{(J2000)} & (cts/ks) & (erg/s) & & (mag)& (mag)&
(mag)& (mag)& (mag)& (mag)& ($M_\odot$) & & class$^d$& \\
\hline
\endfirsthead
\caption{continued.}\\
\hline\hline
Identification$^a$ & RA & DEC & Count rate$^b$ & $\log L_\mathrm{X}^c$& $\log L_\mathrm{X}/L_\mathrm{bol}$& 
 $V$& $R$& $I$& $J$& $H$& $K_\mathrm{s}$& Mass & Sp.T& Disc& Mem.$^e$\\
   &\multicolumn{2}{c}{(J2000)} & (cts/ks) & (erg/s) & & (mag)& (mag)&
(mag)& (mag)& (mag)& (mag)& ($M_\odot$) & & class$^d$& \\
\hline
\endhead
\hline
\endfoot
\hline
\multicolumn{16}{l}{\begin{minipage}{23cm}
{\bf Notes.} 
$^{(a)}$ Stars labelled DM, LOri and LOri-SOC are from \citet{dm99},
\citet{barrado04}, and \citet{bouy09lam}, respectively. The GSC and 2MASS
identifications correspond to the stars labelled c, d, e, g, and Ton~72 in
\citet{murdin77}, respectively.
$^{(b)}$ MOS-equivalent count rates in the 0.3$-$7.8~keV band; upper limits
higher than 2~cts/ks are overestimated due to close bright sources.
$^{(c)}$ X-ray luminosity in the 0.3$-$8~keV band. 
$^{(d)}$ Disc classification from \citet{barrado07} and \citet{hernandez09};
for Class~II or EV stars we also indicate whether there is spectroscopic
evidence for accretion (a) or not (n).
$^{(e)}$ An ``s'' in this column indicates spectroscopically-confirmed
members (i.e. objects with spectroscopic youth features and radial velocity
consistent with membership).
\end{minipage}}\\
\endlastfoot
$\lambda$~Ori\,D       & 05:35:03.02& $+$09:56:04.7& $<  2.61$& $< 29.64$& $< -5.45$&  9.65 & \ldots& \ldots& \ldots& 9.57& 9.49& 2.36& B9   & DD  & -\\
HD 245275              & 05:35:44.86& $+$09:55:24.4& $<  2.70$& $< 29.65$& $< -4.92$& 10.46 &  9.97 &  9.82&  9.80&  9.71&  9.67& 2.54& A7   & DD  & -\\
GSC 00705-00822        & 05:35:09.47& $+$10:00:38.0& $<  0.22$& $< 28.57$& $< -5.48$& 12.16 & 11.77 & 11.42& 10.92& 10.63& 10.56& 1.31& -    & -   & -\\
GSC 00705-00766        & 05:35:19.55& $+$10:01:57.8& $<  0.25$& $< 28.62$& $< -5.41$& 12.62 & 11.89 & 11.23& 10.28&  9.67&  9.49& 1.29& -    & -   & -\\
GSC 00705-00756        & 05:35:13.95& $+$10:02:55.8& $<  0.22$& $< 28.57$& $< -5.37$& 12.46 & 12.02 & 11.64& 11.07& 10.76& 10.66& 1.29& -    & -   & -\\
GSC 00705-00855        & 05:35:26.90& $+$09:59:45.7& $<  0.19$& $< 28.50$& $< -5.43$& 12.79 & 12.11 & 11.50& 10.53& 10.05&  9.85& 1.44& -    & -   & -\\
2MASS J05352828+1004267& 05:35:28.29& $+$10:04:26.7& $<  0.27$& $< 28.65$& $< -4.89$& 13.93 & 13.16 & 12.45& 11.35& 10.72& 10.55& 1.13& -    & -   & -\\
DM 48                  & 05:35:55.86& $+$09:56:21.7& $< 23.57$& $< 30.60$& $< -2.22$& 16.85 & 15.69 & 14.39& 13.01& 12.24& 11.94& 0.38& -    & -   & s\\
LOri 007               & 05:34:29.53& $+$09:48:58.3& $<  0.43$& $< 28.86$& $< -4.57$& 14.92 & 13.72 & 12.78& 11.70& 11.10& 10.89& 0.61& -    & III & -\\
LOri 009               & 05:34:46.35& $+$10:06:35.8& $<  0.37$& $< 28.79$& $< -4.54$& 14.65 & 13.70 & 12.95& 11.84& 11.11& 10.92& 1.01& -    & III & -\\
LOri 011               & 05:34:44.66& $+$09:53:58.0& $<  0.20$& $< 28.52$& $< -4.78$& 15.01 & 13.84 & 13.01& 11.60& 10.78& 10.55& 0.84& -    & III & -\\
LOri 012               & 05:35:05.96& $+$09:52:08.0& $<  0.17$& $< 28.45$& $< -4.85$& 14.83 & 13.80 & 13.03& 11.82& 10.97& 10.79& 0.98& -    & III & -\\
LOri 015               & 05:34:21.82& $+$10:04:14.9& $<  0.49$& $< 28.91$& $< -4.38$& 14.83 & 13.83 & 13.05& 11.87& 11.13& 10.91& 0.96& -    & III & -\\
LOri 027               & 05:35:11.01& $+$10:07:36.4& $<  0.37$& $< 28.79$& $< -4.36$& 15.72 & 14.49 & 13.50& 12.38& 11.72& 11.50& 0.57& -    & III & -\\
LOri 030               & 05:35:12.54& $+$09:55:19.5& $<  0.35$& $< 28.77$& $< -4.30$& \ldots& 14.95 & 13.74& 12.43& 11.69& 11.43& 0.41& -    & III & s\\
LOri 053               & 05:34:36.73& $+$09:52:58.3& $<  0.32$& $< 28.72$& $< -3.98$& 17.73 & 16.08 & 14.72& 13.17& 12.52& 12.29& 0.36& -    & III & s\\
LOri 063               & 05:35:19.14& $+$09:54:42.4& $< 60.99$& $< 31.01$& $< -1.45$& \ldots& 16.80 & 15.34& 13.76& 13.07& 12.66& 0.30& M4.5 & II a& s\\
LOri 069               & 05:34:43.97& $+$09:48:35.6& $<  0.36$& $< 28.78$& $< -3.79$& \ldots& 16.89 & 15.20& 13.38& 12.77& 12.42& 0.27& -    & III & s\\
LOri 073               & 05:34:46.82& $+$09:50:37.9& $<  0.46$& $< 28.89$& $< -3.62$& \ldots& 16.84 & 15.28& 13.64& 12.99& 12.71& 0.29& M5.0 & III & s\\
LOri 076               & 05:35:10.96& $+$09:57:43.8& $<  0.23$& $< 28.58$& $< -4.02$& \ldots& 17.39 & 15.81& 14.22& 13.53& 13.20& 0.26& -    & III & s\\
LOri 079               & 05:34:48.26& $+$09:59:53.9& $<  0.20$& $< 28.52$& $< -3.69$& \ldots& 17.51 & 16.00& 14.22& 13.54& 13.34& 0.27& -    & EV n& s\\
LOri 082               & 05:36:00.81& $+$09:52:57.1& $<  0.65$& $< 29.04$& $< -3.17$& \ldots& 17.57 & 16.02& 14.20& 13.57& 13.28& 0.23& M4.5 & III & s\\
LOri 085               & 05:35:21.52& $+$09:53:29.2& $<  0.22$& $< 28.57$& $< -3.64$& \ldots& 17.65 & 16.04& 14.19& 13.62& 13.23& 0.19& -    & II  & s\\
LOri 086               & 05:34:11.58& $+$09:49:15.0& $<  0.62$& $< 29.02$& $< -3.15$& \ldots& 17.59 & 16.09& 14.48& 13.87& 13.50& 0.27& -    & III & s\\
LOri 087               & 05:34:33.77& $+$09:55:34.2& $<  0.30$& $< 28.70$& $< -3.46$& \ldots& 17.54 & 16.09& 14.19& 13.60& 13.28& 0.31& M4.5 & EV n& s\\
LOri 088               & 05:34:49.50& $+$09:58:46.8& $<  0.31$& $< 28.71$& $< -3.50$& \ldots& 17.78 & 16.10& 14.14& 13.54& 13.23& 0.15& -    & III & s\\
LOri 089               & 05:35:04.00& $+$10:07:26.8& $<  0.35$& $< 28.77$& $< -3.41$& \ldots& 17.79 & 16.15& 14.38& 13.84& 13.51& 0.16& M5.0 & EV n& -\\
LOri 091               & 05:34:35.86& $+$09:54:26.5& $<  0.31$& $< 28.71$& $< -3.50$& \ldots& 18.01 & 16.18& 14.18& 13.56& 13.29& 0.16& M5.5 & EV n& s\\
LOri 092               & 05:35:50.97& $+$09:51:03.5& $<  0.32$& $< 28.73$& $< -3.43$& \ldots& 17.84 & 16.19& 14.44& 13.84& 13.54& 0.15& -    & III & s\\
LOri 093               & 05:34:41.20& $+$09:50:16.3& $<  0.38$& $< 28.80$& $< -3.35$& \ldots& 17.82 & 16.21& 14.46& 13.84& 13.60& 0.17& -    & III & s\\
LOri 094               & 05:34:43.17& $+$10:01:59.8& $<  0.28$& $< 28.67$& $< -3.48$& \ldots& 18.03 & 16.28& 14.40& 13.80& 13.42& 0.11& M5.5 & III & s\\
LOri 095               & 05:35:24.18& $+$09:55:15.4& $<  0.24$& $< 28.60$& $< -3.49$& \ldots& 17.96 & 16.35& 14.56& 13.91& 13.61& 0.16& M6.0 & III & s\\
LOri 096               & 05:35:11.13& $+$09:57:19.6& $<  0.42$& $< 28.85$& $< -3.24$& \ldots& 18.02 & 16.37& 14.63& 13.96& 13.64& 0.14& -    & II n& s\\
LOri 099               & 05:34:45.59& $+$10:05:48.9& $<  0.37$& $< 28.79$& $< -3.30$& \ldots& 18.14 & 16.42& 14.71& 14.07& 13.68& 0.11& M5.5 & III & -\\
LOri 100               & 05:35:00.10& $+$09:46:14.0& $<  0.52$& $< 28.94$& $< -3.13$& \ldots& 18.08 & 16.43& 14.77& 14.04& 13.82& 0.14& M5.5 & III & s\\
LOri 102               & 05:35:22.02& $+$09:52:52.3& $<  0.24$& $< 28.60$& $< -3.46$& \ldots& 18.24 & 16.50& 14.63& 14.08& 13.81& 0.10& -    & III & s\\
LOri 103               & 05:35:22.56& $+$09:45:01.9& $<  0.37$& $< 28.79$& $< -3.25$& \ldots& 18.30 & 16.55& 14.64& 14.13& 13.83& 0.10& -    & EV  & s\\
LOri 104               & 05:35:07.07& $+$09:54:01.5& $<  0.29$& $< 28.69$& $< -3.30$& \ldots& 18.48 & 16.71& 14.67& 14.14& 13.72& 0.08& -    & II  & s\\
LOri 105               & 05:34:17.58& $+$09:52:29.7& $<  3.70$& $< 29.79$& $< -2.20$& \ldots& 18.58 & 16.75& 14.92& 14.34& 13.99& 0.17& -    & III & s\\
LOri 106               & 05:35:28.77& $+$09:54:10.2& $<  0.19$& $< 28.50$& $< -3.53$& \ldots& 18.48 & 16.76& 14.78& 14.16& 13.74& 0.12& M5.5 & II a& s\\
LOri 107               & 05:35:55.19& $+$09:52:20.0& $<  0.90$& $< 29.18$& $< -2.89$& \ldots& 18.85 & 16.78& 14.66& 13.99& 13.62& 0.10& M6.0 & III & s\\
LOri 108               & 05:35:26.04& $+$10:08:09.8& $<  0.93$& $< 29.19$& $< -2.81$& \ldots& 18.64 & 16.80& 14.84& 14.26& 13.92& 0.13& -    & III & -\\
LOri 109               & 05:34:08.54& $+$09:50:43.5& $<  0.67$& $< 29.05$& $< -2.92$& \ldots& 18.67 & 16.81& 14.96& 14.47& 14.18& 0.16& M5.5 & III & s\\
LOri 112               & 05:34:33.53& $+$09:43:55.5& $<  0.65$& $< 29.04$& $< -2.91$& \ldots& 18.72 & 16.87& 14.99& 14.36& 14.15& 0.14& -    & III & -\\
LOri 115               & 05:34:46.32& $+$10:02:31.9& $<  8.81$& $< 30.17$& $< -1.63$& \ldots& 18.80 & 17.08& 15.45& 14.82& 14.59& 0.09& M5.0 & EV n& -\\
LOri 116               & 05:35:12.03& $+$10:01:04.3& $<  0.20$& $< 28.52$& $< -3.30$& \ldots& 19.05 & 17.17& 15.34& 14.57& 14.41& 0.15& M5.5 & EV n& s\\
LOri 117               & 05:35:07.95& $+$10:00:06.2& $<  0.23$& $< 28.58$& $< -3.31$& \ldots& 19.24 & 17.21& 15.10& 14.36& 14.17& 0.09& M6.0 & EV n& -\\
LOri 118               & 05:35:24.41& $+$09:53:51.9& $<  0.21$& $< 28.55$& $< -3.28$& \ldots& 19.10 & 17.23& 15.27& 14.69& 14.18& 0.11& M5.5 & II n& s\\
LOri 120               & 05:34:46.21& $+$09:55:37.7& $<  0.22$& $< 28.57$& $< -3.23$& \ldots& 19.23 & 17.34& 15.33& 14.77& 14.34& 0.10& M5.5 & II n& s\\
LOri 122               & 05:34:35.44& $+$09:51:18.6& $<  0.29$& $< 28.68$& $< -3.09$& \ldots& 19.31 & 17.38& 15.43& 14.85& 14.46& 0.11& -    & EV  & s\\
LOri 123               & 05:34:20.47& $+$10:05:22.4& $<  0.64$& $< 29.03$& $< -2.75$& \ldots& 19.53 & 17.42& \ldots& \ldots& \ldots& 0.14& - & -   & -\\
LOri 124               & 05:34:14.25& $+$09:48:26.3& $<  0.61$& $< 29.01$& $< -2.68$& \ldots& 19.30 & 17.45& 15.66& 15.06& 14.78& 0.15& M5.5 & III & s\\
LOri 126               & 05:35:39.85& $+$09:53:24.1& $<  0.26$& $< 28.64$& $< -3.06$& \ldots& 19.52 & 17.52& 15.62& 15.04& 14.67& 0.12& M6.5 & II n& s\\
LOri 128               & 05:35:06.31& $+$09:58:01.3& $<  0.22$& $< 28.57$& $< -3.12$& \ldots& 19.53 & 17.58& 15.62& 15.10& 14.77& 0.10& -    & III & -\\
LOri 130               & 05:34:56.54& $+$09:42:33.0& $<  0.50$& $< 28.92$& $< -2.73$& \ldots& 19.44 & 17.63& 15.73& 15.26& 14.73& 0.11& M5.5 & III & s\\
LOri 132               & 05:34:29.16& $+$09:47:07.2& $<  0.44$& $< 28.87$& $< -2.83$& \ldots& 19.99 & 17.82& 15.58& 14.96& 14.91& 0.07& -    & EV  & -\\
LOri 134               & 05:35:22.85& $+$09:55:07.1& $<  0.27$& $< 28.65$& $< -3.07$& \ldots& 19.91 & 17.90& 15.54& 14.94& 14.67& 0.06& M5.0 & EV n& -\\
LOri 135               & 05:35:09.33& $+$09:52:44.2& $<  0.18$& $< 28.48$& $< -3.18$& \ldots& 19.91 & 17.90& 15.67& 15.08& 14.91& 0.07& M7.0 & III & s\\
LOri 136               & 05:34:38.37& $+$09:58:11.6& $<  0.23$& $< 28.58$& $< -3.13$& \ldots& 20.06 & 17.92& 15.56& 14.83& 14.58& 0.06& M5.0 & EV  & -\\
LOri 139               & 05:35:44.34& $+$10:05:54.4& $<  0.50$& $< 28.92$& $< -2.55$& \ldots& 20.04 & 18.16& 16.16& 15.53& 15.06& 0.09& M6.0 & II n& -\\
LOri 140               & 05:34:19.27& $+$09:48:27.5& $<  0.61$& $< 29.01$& $< -2.53$& \ldots& 20.34 & 18.21& 15.98& 15.22& 14.75& 0.07& M7.0 & II a& s\\
LOri 143               & 05:35:00.95& $+$09:58:20.3& $<  0.21$& $< 28.55$& $< -2.93$& \ldots& 20.32 & 18.30& 16.11& 15.61& 15.23& 0.07& M6.5 & III & s\\
LOri 146               & 05:35:00.16& $+$09:52:40.9& $<  0.29$& $< 28.69$& $< -2.76$& \ldots& 20.88 & 18.60& 16.23& 15.47& 14.94& 0.06& -    & EV  & s\\
LOri 147               & 05:35:06.26& $+$09:46:53.8& $<  0.27$& $< 28.65$& $< -2.65$& \ldots& 20.54 & 18.60& 16.58& 15.93& 15.62& 0.08& M5.5 & EV n& -\\
LOri 150               & 05:35:07.50& $+$09:49:32.9& $<  0.23$& $< 28.58$& $< -2.70$& \ldots& 21.29 & 19.00& 16.66& 16.13& 15.56& 0.06& M8.0 & III & -\\
LOri 154               & 05:34:19.81& $+$09:54:20.6& $<  0.38$& $< 28.80$& $< -2.43$& \ldots& 21.79 & 19.31& 16.80& 16.14& 15.51& 0.05& M8.0 & -   & -\\
LOri 156               & 05:34:36.28& $+$09:55:32.2& $<  0.27$& $< 28.65$& $< -2.48$& \ldots& 22.05 & 19.59& 17.06& 16.34& 15.89& 0.05& M8.0 & II a& -\\
LOri 162               & 05:35:04.44& $+$09:57:33.6& $<  0.29$& $< 28.69$& $< -2.22$& \ldots& 23.22 & 20.42& 17.64& 16.90& 16.52& 0.04& -    & -   & -\\
LOri 163               & 05:35:18.36& $+$09:56:52.8& $<  0.58$& $< 28.99$& $< -1.82$& \ldots& 22.96 & 20.42& 17.86& 17.02& 16.76& 0.03& -    & -   & -\\
LOri 167               & 05:35:14.19& $+$09:54:07.5& $<  0.30$& $< 28.70$& $< -2.14$& \ldots& 23.86 & 20.90& 17.88& 17.15& 16.62& 0.03& M9-L2& -   & -\\
LOri 167B              & 05:35:14.51& $+$09:54:07.4& $<  0.21$& $< 28.55$& $< -1.36$& \ldots& \ldots& 23.23& 20.19& 19.41& 18.31& 0.01& L6   & -   & -\\
LOri-SOC-3             & 05:35:00.90& $+$09:54:40.4& $<  0.22$& $< 28.55$& $< -2.16$& \ldots& \ldots& 20.84& 18.12& 17.58& 16.89& 0.03& -    & -   & -\\
LOri-SOC-5             & 05:35:03.03& $+$09:55:47.3& $<  0.25$& $< 28.60$& $< -1.93$& \ldots& \ldots& 22.08& 18.73& 18.36& 17.25& \ldots& -  & -   & -\\
LOri-SOC-6             & 05:35:03.72& $+$09:54:13.9& $<  0.23$& $< 28.56$& $< -3.15$& \ldots& \ldots& 17.33& 15.65& 15.14& 14.43& 0.20& -    & -   & -\\
LOri-SOC-10            & 05:35:07.79& $+$09:55:21.8& $<  0.38$& $< 28.78$& $< -2.44$& \ldots& \ldots& 19.80& 16.89& 16.32& 15.63& 0.03& -    & -   & -\\
LOri-SOC-11            & 05:35:09.11& $+$09:54:36.1& $<  0.31$& $< 28.69$& $< -2.68$& \ldots& \ldots& 19.00& 16.46& 16.00& 15.36& 0.04& -    & -   & -\\
LOri-SOC-12            & 05:35:10.65& $+$09:57:23.3& $<  0.60$& $< 29.00$& $< -1.52$& \ldots& \ldots& 21.62& 18.66& 17.98& 17.12& 0.03& -    & -   & -\\
LOri-SOC-13            & 05:35:10.78& $+$09:56:06.5& $<  0.41$& $< 28.83$& $< -2.72$& \ldots& \ldots& 18.16& 15.94& 15.56& 14.92& 0.07& -    & -   & -\\
\end{longtable}
\end{landscape}
}

\end{document}